# Switchable Magnetoelectric Transport in Graphene via a Van der Waals Multiferroic

Miuko Tanaka*[1], Shunta Aoki[1], Ikoi Sato[1], Hao Ou[2], Itishree Pradhan[1], Ngoc Han Tu[5], Yangsong Chen[1], Tomohiro Ishii[1], Kenji Watanabe[3], Takashi Taniguchi[4], Michihisa Yamamoto[5,6], Masayuki Hashisaka[1], Jiang Pu[2], Naoki Ogawa[5], Toshiya Ideue*[1]

*[1]Institute for Solid State Physics, The University of Tokyo, Kashiwa-shi, Japan*

*[2] Department of Physics, Institute of Science Tokyo, Meguro-ku, Japan*

*[3]Research Center for Electronic and Optical Materials, National Institute for Materials Science, Tsukuba, Japan*

*[4]Research Center for Materials Nanoarchitectonics, National Institute for Materials Science, Tsukuba, Japan*

*[5]Center for Emergent Matter Science, RIKEN, Wako-shi, Japan*

*[6]Quantum-Phase Electronics Center and Department of Applied Physics, The University of Tokyo, Tokyo, Japan*

*Corresponding authors

## Abstract

Electric and magnetic control of transport properties at atomic interfaces is central to the development of next-generation electronics and spintronics. Van der Waals multiferroics—materials that simultaneously host dielectric and magnetic orders down to the monolayer limit—offer a promising platform for such interfacial control, yet the realization of electronic functionalities that exploit the unique attributes of van der Waals multiferroics has largely remained elusive.

Here, we realize a van der Waals heterostructure comprising graphene and the multiferroic $CuCrP_2S_6$, enabling gate-switchable magnetoelectric transport in graphene, mediated by the multiferroic layer. The charge-neutrality resistance peak of graphene exhibits pronounced hysteresis arising from polarization flip in the multiferroic state. Application of an in-plane magnetic field shifts this peak in a polarization-dependent manner, revealing magnetic-field-induced polarization modulation—a direct signature of the magnetoelectric effect. Furthermore, cooling the device under an applied electric field enables domain control of the multiferroic order, allowing reversible switching of the interfacial magnetoelectric transport.

These results provide the first demonstration of interfacial magnetoelectric transport in a vdW heterostructure, and establish a pathway for engineering two-dimensional van der Waals interfaces for functional device applications.

## Introduction

Solid interfaces often host physical properties and functionalities that are absent in the constituent materials [1-5]. Among them, van der Waals interfaces offer a particularly powerful platform, as arbitrary combinations of crystals with different properties can be assembled with precise control over stacking sequence and alignment. This flexibility has enabled the discovery of a wide range of emergent interfacial phenomena in van der Waals heterostructures, including those at magnet/semiconductor [6-9, 18, 19], ferroelectric/semiconductor [10-17], and magnet/superconductor interfaces [20, 21].

The recent discovery of van der Waals multiferroics—materials that combine dielectric and magnetic orders down to the monolayer limit—opens further opportunities to design interfaces with coupled electronic, magnetic, and dielectric degrees of freedom [22-32]. In particular, heterostructures that integrate van der Waals multiferroics with two-dimensional quantum materials such as graphene promise interfacial magnetoelectric transport with dual tunability by electric and magnetic fields, a functionality that is difficult to achieve in conventional bulk multiferroics [33-46]. Despite intensive studies revealing exotic optical and electrical responses of thin van der Waals multiferroics associated with their magnetoelectric nature [22-32, 42-46], demonstrations of transport functionalities directly reflecting this interfacial electromagnetism have remained elusive.

Here, we demonstrate gate-tunable magnetoelectric transport in graphene by exploiting an interface with multiferroic $CuCrP_2S_6$ (CCPS) (Fig. 1a). The resistance of graphene is effectively modulated by dielectric and magnetic states of CCPS. The resistance peak corresponding to the charge neutrality point (CNP) of graphene exhibits pronounced hysteresis, reflecting the polarization switching of CCPS. The CNP position depends strongly on the spin configuration of CCPS, and shifts under applied magnetic fields, providing clear evidence that transport is governed by magnetoelectric coupling. This magnetic-field-dependent shift of CNP is suppressed above the magnetic transition temperature of CCPS, confirming its interfacial magnetoelectric origin. Moreover, by cooling the device under an applied electric field, we control multiferroic domain configurations and thereby achieve reversible switching of the magnetoelectric transport.

This work establishes the first demonstration of interfacial magnetoelectric transport in a vdW heterostructure. It further provides a design principle for functional van der Waals interfaces with emergent magnetoelectric properties, applicable to a broad range of material combinations and offering a pathway to advanced electronic and spintronic functionalities.

**Transport signatures of polarization flip in graphene/multiferroic interfaces**

CCPS is an air-stable van der Waals multiferroic insulator. At room temperature, $Cu^+$ ions are randomly distributed along the out-of-plane direction around centrosymmetric positions (Fig. 1b, left panel; see SI Section 1 for the crystal structure). Under an applied electric field, the center of this distribution shifts out of the plane with slow dynamics (Fig. 1b, right panel), resulting in ferroelectric-like hysteresis behavior, as previously reported and confirmed by our piezoresponse force microscopy measurements (see SI Section 3) [22-25, 45].

At 190 K, the $Cu^+$ ions start to freeze, and they settle into an ordered configuration at 145 K, marking a structural phase transition to the non-centrosymmetric space group Pc (Fig. 1c; see SI Section 1) [26-28, 32, 42, 43, 45]. In this phase, neighboring $Cu^+$ ions are displaced alternately in opposite out-of-plane directions. Importantly, the transition is accompanied by a distortion of the $CrS_6$–$P_2S_6$ framework that breaks inversion symmetry; the two oppositely displaced $Cu^+$ ions therefore occupy inequivalent sites, and complete cancellation of their dipole moments is not enforced by symmetry. Upon further cooling, the Cr spins, arranged in a honeycomb network within each van der Waals layer, order antiferromagnetically below the Néel temperature ($T_N \approx 32$ K), giving rise to a multiferroic state in which dielectric and magnetic orders are coupled (Fig. 1c).

In this study, we use CCPS as a gate dielectric to control the carrier density of graphene, realizing interfacial magnetoelectric transport. Since the graphene placed on top is primarily sensitive to the electrostatic potential at the CCPS surface, it directly probes the switching of the surface polarization. We fabricate two different types of devices composed of CCPS, monolayer graphene, and hexagonal boron nitride (hBN), stacked on a pre-patterned back gate (Figs. 1a, d, e). The first device has the structure back gate / CCPS / hBN (6~8 nm) / graphene (Fig. 1d), and the second back gate / CCPS / graphene (Fig. 1e). We call these W-BN (with hBN) and WO-BN (without hBN), respectively. Figures 1f and 1g compare the graphene transfer curves of the two devices at $T$ = 2 K. Both devices exhibit clear hysteresis reflecting the hysteretic polar behavior of CCPS. We note that the polarization of the entire bulk does not flip under experimentally accessible electric fields: theoretical calculations indicate that the barrier for the $Cu^+$ rearrangement grows rapidly with the flipping layer number [26], and our capacitance measurements directly show that the bulk-averaged dielectric constant of the flake remains unchanged under gating while the surface polarization is reversed (see SI Section 5). These observations are inconsistent with a bulk-uniform polarization flip and strongly suggest that the polarization switching discussed here is confined to the near-surface region (Fig. 1c). Given the underlying nominally antiferroelectric (AFE) bulk ground state, it is reasonable to consider ferrielectric surface states, in which the two $Cu^+$ displacements do not fully cancel and a finite net polarization remains [26, 43, 44]. Throughout this work, we label these gate-switchable surface polar states with positive (negative) net polarization as FiE1 (FiE2); the two states are interchanged by the gate electric field but, owing to the broken inversion symmetry of the framework, they are not required to be equivalent (see SI Section 6 for the detailed picture).

An important observation is that the direction of the CNP shift with respect to the sweep direction is opposite between the two devices. This contrasting behavior can be understood by decomposing the graphene carrier density into four contributions: (i) electrostatic doping proportional to $V_G$, (ii) electrostatic doping from the CCPS surface spontaneous polarization, (iii) surface-polarization-dependent charge transfer at the CCPS/graphene interface, and (iv) surface-polarization-independent offset doping at the CCPS/graphene interface. In the W-BN device, the hBN spacer suppresses (iii) and (iv), and the electrostatic contributions (i) and (ii) dominate, whereas in the WO-BN device all four contributions coexist, as detailed below (see SI Section 7 for schematics of how the four contributions combine at each step of the switching cycle).

In W-BN device, thin hBN is inserted between graphene and CCPS, which can be regarded as a parallel-plate capacitor composed of a back gate and graphene, with CCPS and thin hBN acting as the gate dielectrics. The carrier density in graphene $n$ is modulated by both the gate voltage $V_G$ and the effective spontaneous polarization P of CCPS, defined as a sheet polarization (in units of $C/m^2$), according to the following expression (see SI Section 13 for the derivation and detailed model).

$$n = \frac{V_{\mathrm{G}} + \frac{d_{CCPS}}{\varepsilon_{CCPS}} P}{\mathrm{e}\left(\frac{d_{CCPS}}{\varepsilon_{CCPS}} + \frac{d_{hBN}}{\varepsilon_{hBN}}\right)} + \mathrm{const.} \qquad (1)$$

Here, $\varepsilon_{CCPS} \cong 7.7\varepsilon_0$ and $\varepsilon_{hBN} \cong 3.8\varepsilon_0$ denote the dielectric constant of CCPS and hBN, respectively, while $d_{CCPS}$ and $d_{hBN}$ are their thicknesses (see SI Section 5 for the measurement of $\varepsilon_{CCPS}$). The constant offset term accounts for the intrinsic doping of graphene, which is known to be less than $2\times10^{11}$ $cm^{-2}$ at hBN/graphene interfaces [67, 68] and is thus smaller than the first term. Using Eq. 1, the CNP gate voltages in the FiE1 and FiE2 states, $V_{G,\ CNP,\ FiE1}$ and $V_{G,\ CNP,\ FiE2}$, are obtained as follows:

$$V_{\mathrm{G,CNP,FiE1(2)}} = -\frac{d_{CCPS}}{\varepsilon_{CCPS}} P_{\mathrm{FiE1\,(2)}} + \mathrm{const.}, \quad (2)$$

where $P_{\mathrm{FiE1\,(2)}}$ denotes the effective spontaneous surface polarization of the FiE1 (FiE2) state. When the gate voltage is swept toward positive values, CCPS adopts the FiE1 state with positive surface polarization (positive bound charges on the top side), which acts on graphene as an effective positive gate-voltage offset in the W-BN device, shifting the CNP toward negative gate voltages (blue curve in Fig. 1f). Conversely, when the gate voltage is swept toward negative values, CCPS switches to the FiE2 state with negative surface polarization, producing an effective negative gate-voltage offset and shifting the CNP toward positive gate voltages (red curve in Fig. 1f). Owing to the non-volatile ferrielectric nature of CCPS, the CNP appears at distinct $V_G$ with opposite sign for the two sweep directions, indicating two polarization states with opposite signs of spontaneous polarizations. The coercive electric field, at which the polarization hysteresis begins to open, is estimated to be 0.17–0.21 V/nm, which is comparable to the coercive field obtained from piezoresponse force microscopy measurements (see SI Sections 3 and 15). Applying a sufficiently large gate voltage more than 11 V ensures the full saturation of polarization (see SI Section 15).

The polarization magnitude can be quantified from the CNP peak positions for the two sweep directions. The observed CNP separation of approximately 6 V corresponds to an effective polarization difference of 1.05 $\mu C/cm^2$ between the two surface polarization states, which is smaller than the bulk polarization values of 2.5–16 $\mu C/cm^2$ reported at room temperature [22, 24], although the substantial variability among these earlier reports renders a direct comparison difficult. This reduction can be attributed to the surface confinement of the polarization flip or to a difference in the Cu-ion configurations between room temperature and low temperature.

In contrast, in the WO-BN device, CCPS is directly attached to graphene (Fig. 1e), and the transfer curve exhibits a reversed behavior, so-called anti-hysteresis, in which the resistance peak shifts in the same direction as the pre-applied gate voltage (Fig. 1g). This behavior cannot be explained by the electrostatic contributions (i) and (ii) alone, and indicates that the polarization dependent interfacial charge transfer (iii) is at work (see SI Section 7 for the detailed schematics, and SI Section 9 for a quantification of the carrier density and its hysteresis by Hall and quantum-oscillation measurements). Hysteresis caused by slow charge trapping — for example, by molecules such as water or organics adsorbed at the interface — is known to depend on the gate-sweep rate, with time constants of seconds to minutes [11, 13-15]. In our measurements, no sweep-rate dependence is observed over this range (see SI Section 14), which excludes slow interfacial trapping as the origin of the anti-hysteresis. The anti-hysteresis is instead consistent with a fast charge transfer whose magnitude depends on the polarization state of the CCPS surface. When the polarization is positive (positive bound charge on the top surface), holes are transferred to graphene, shifting the CNP in the same direction as the pre-applied gate voltage, and vice versa (Fig. 1g). The magnitude of this charge transfer is reproducible across samples, indicating an origin intrinsic to the CCPS polar states (see SI Section 8 for comparison across samples). In addition, the shift of the center of the two CNP peaks indicates that a significant, gate-unswitchable offset doping (iv) is present at the interface (see SI Sections 7 and 9 for the detailed nature of this contribution). When we consider total four contributions (i)~(iv), the carrier density and CNP of graphene are expressed as [12]:

$$n = \frac{\varepsilon_{CCPS}}{\mathrm{e}d_{CCPS}} V_{\mathrm{G}} + \frac{1}{\mathrm{e}} P - \Delta n_{\mathrm{FiE1\,(2)}} + \mathrm{const.} \quad (3)$$

$$V_{\mathrm{CNP,FiE1\,(2)}} = -\frac{d_{CCPS}}{\varepsilon_{CCPS}} \left(\left[P_{\mathrm{FiE1\,(2)}} - \mathrm{e}\Delta n_{\mathrm{FiE1\,(2)}}\right]\right) + \mathrm{const.}, \quad (4)$$

where $\Delta n_{\mathrm{FiE1\,(FiE2)}}$ denotes the charge transferred to graphene in the FiE1 (FiE2) state. Based on the anti-hysteretic behavior—opposite to that in W-BN device—the CNP position in WO-BN device is inferred to be dominated by the charge-transfer term, that is, $|P_{\mathrm{FiE1}} - P_{\mathrm{FiE2}}|/\mathrm{e} < |\Delta n_{\mathrm{FiE1}} - \Delta n_{\mathrm{FiE2}}|$. Here, correspondence between the two CNP peaks and the polar states is reversed between the two device structures: the FiE1 peak appears on the negative-$V_G$ side of the hysteresis in the W-BN device, but on the positive-$V_G$ side in the WO-BN device (see SI Section 7). Although the two device structures exhibit opposite hysteresis behavior, both demonstrate that the interfacial graphene transport is modulated by the surface polar states of CCPS, thereby providing an effective probe of the polarization switching.

**Magnetoelectric Control of Graphene Transport**

We next investigate the effect of a magnetic field on the graphene transport in the two device structures. In both polar states FiE1 and FiE2, the CNP of graphene exhibits a clear shift under an in-plane magnetic field (Figs. 2a-d). Because the CNP of monolayer graphene itself does not shift under a magnetic field [47], this magnetic-field-dependent shift indicates that the surface polarization of multiferroic CCPS is modulated by the magnetic field—i.e., a magnetoelectric (ME) effect. These results provide a direct and quantitative observation of the ME effect in micro-scale multiferroic samples via transport measurements. The orientation of the magnetic field with respect to the crystal axes is provided in SI Section 10; no significant dependence on the *B* direction is observed. The ME effect is quantitatively comparable among different samples for the same polar states (see SI Section 8).

A closer look at Figs. 2c and 2d reveals that the magnitude of the shift clearly differs between the two polar states. The CNP shift of the FiE1 peak is larger than that of the FiE2 peak, in both the W-BN and WO-BN devices, and the relative ratio of the shifts is comparable between the two device structures (note that the FiE1 peak appears at negative $V_G$ in the W-BN device but at positive $V_G$ in the WO-BN device, as explained in the previous section). The ME response thus clearly distinguishes the two polar states, demonstrating that FiE1 and FiE2 are indeed inequivalent, as anticipated from the inversion-symmetry-broken $CrS_6$–$P_2S_6$ framework. Figure 2e schematically summarizes the polarization loop and the two surface states. We note that the polarization loops in Figs. 2e and 4a are drawn with different magnitudes of the positive and negative polarizations; this asymmetry between FiE1 and FiE2 is established by the field-cooling experiments discussed in the final section (see also SI Sections 6 and 9) and does not affect the discussion here.

The saturation field $B_s$, above which $V_{G,\ CNP}$ becomes nearly independent of $B$, corresponds to the transition from the antiferromagnetic to the forced ferromagnetic state, in which the spins are aligned parallel to the applied magnetic field. We also note that $B_s$ appears slightly smaller in the FiE2 state than in the FiE1 state in the W-BN device (Fig. 2c), which might reflect a polarization-state dependence of the interlayer exchange coupling [26]; this difference is, however, less pronounced in the WO-BN device (see SI Section 18 for detailed discussion).

Importantly, for each of the FiE1 and FiE2 states, the sign of the CNP peak shift under a magnetic field is the same in the two device structures, with comparable shift magnitudes (Figs. 2c, d). This indicates that the interfacial charge-transfer term (iii) does not exhibit a significant $B$ dependence, and that the peak shift is dominated by the $B$ dependence of the polarization term (ii) in Eqs. 2 and 4, which contribute to graphene with the same sign in two device structures. Note that the $B$ dependence of the dielectric constant is too small to account for the observed behavior (see SI Section 5 for dielectric constant measurement).

Although the magnetic field modulation of CNP appears to be primarily an even function of the magnetic field (second-order ME effect), we also find a $B$-odd component in FiE1 state. Figure 2f presents the magnetic field dependence of the CNP position ($V_{G,\ CNP}$) for opposite sweep directions of the magnetic field in FiE1 state of WO-BN device. A clear hysteresis is observed between the two sweep directions, corresponding to the shift in x($B$)-axis of approximately 0.4 T. This large shift cannot be attributed to instrumental artifacts such as remanent magnetization of the superconducting magnet or measurement setup, thereby indicating the presence of an intrinsic $B$-odd contribution, i.e., linear ME effect. Note that a $B$-odd response cannot arise in pristine graphene owing to Onsager's reciprocal theorem, even if the applied magnetic field is misaligned from the plane, because the measurement is performed in a two-terminal geometry across the gated region.

To isolate the linear ME component, we compute the difference in $V_{G,\ CNP}$ ($B$) between the forward and reverse magnetic-field sweeps ($\Delta V_{G,\ CNP}$) as shown in Fig. 2f bottom panel. The resulting antisymmetric signal exhibits a linear dependence on $B$ and drops down to zero above $B_s$. This behavior suggests that the sign of the linear ME effect is determined by the orientation of Néel vector of the antiferromagnetic ordering, for example via a bilinear coupling of the form $\vec{P} \propto \vec{B} \times \vec{A}$, where the Néel vector $\vec{A}$ is reversed by applying an opposite magnetic field exceeding $B_s$ (Fig. 2g). This is consistent with the symmetry-allowed linear ME effect term in CCPS (see SI Section 12 for detail). Furthermore, the amplitude of this signal decreases with increasing temperature, confirming its origin in the antiferromagnetic order (see SI Section 11). It is emphasized that nonreciprocal interfacial charge modulation is realized through this linear ME effect of CCPS.

Here, the magnitude of the ME effect can be quantitatively evaluated in the same manner as the electric polarization at zero magnetic field. The magnitude of the observed polarization change induced by the linear ME effect reaches approximately 0.011 $\mu C/cm^2$, corresponding to about 1 % of the polarization difference between FiE1 and FiE2 states. Notably, this value is three orders of magnitude larger than previously reported values in bulk CCPS [43]. The discrepancy between our measurements on nanoflakes and the prior report on bulk CCPS may arise from several factors. One possible explanation is the presence of multiple AFM domains—either lateral or vertical—in bulk crystals, which could lead to domain averaging and a suppression of the net ME response. In contrast, our micro-scale devices with typical lateral dimensions of ~2 μm × 6 μm likely contain fewer lateral domains. Furthermore, the graphene probes the surface of the CCPS flake, which reduces cancellation of the ME signal by vertical domains (in the analogous material $CuInP_2S_6$, a typical domain size of ≈ 200 nm has been reported [66]).

Surface effects in nanoflakes may also play an important role via two possible mechanisms. First, because the symmetry at the surface is lower than in the bulk, additional components of the ME tensor can become symmetry-allowed at the surface [54]. As a result, the ME response near the surface may differ from that in the bulk. Although it is difficult to determine from the present data whether this effect enhances or reduces the overall ME response, it is not unexpected that the surface ME contribution could be comparable to or even larger than the bulk contribution. Second, the FiE state is stabilized near the surface or in thin flakes, but not in bulk CCPS. Therefore, the polarization configuration of the ground state near the surface may differ from that in bulk crystals, which could modify the magnitude of the ME effect.

The linear ME effect in the FiE1 state is consistently observed in another WO-BN sample (S3) and a W-BN sample (S7) (see SI Section 11 for the data). In the FiE2 state, a similar component is resolved only in part of the data sets, owing to the poorer signal-to-noise ratio (see SI Section 11).

**Temperature dependence of magnetoelectric transport**

Figures 3a-c show the shift of the CNP peak in FiE1' state (the difference between FiE1 and FiE1' is explained in the next section) at different temperatures across the Néel temperature 32 K. At 2 K and 15 K, the peak shift is saturated at $B_s$=6~6.5 T, corresponding to the transition from AFM to the forced ferromagnetic state where spins are aligned parallel to applied magnetic field (Figs. 3a, b). In contrast, at 40 K, the shift is much smaller and no saturation behavior is observed, indicating the absence of the antiferromagnetic order (Fig. 3c). The kink at $B_s$, which is associated with the antiferromagnetic order, is characterized by taking subtraction of $\Delta V_G/\Delta B$ above and below $B_s$ (Fig. 3d). The magnitude of the kink decreases as temperature increase and vanishes around 30 K, corresponding to the antiferromagnetic to paramagnetic phase transition (Fig. 3e). This temperature is comparable with reported values of Néel temperature, as well as to the anomaly observed in out-of-plane transport measurement (see SI Section 16).

Although long-range antiferromagnetic order disappears above 32 K, the magnetization still varies with the applied magnetic field as the system evolves from a paramagnetic state toward a field-polarized state. Owing to this field-induced change in magnetization, the ME effect does not completely vanish above the Néel temperature. Similar behavior has also been observed in other multiferroic materials above the magnetic transition temperature [51, 52].

**Switchable magnetoelectric transport via multiferroic domain control**

Finally, we demonstrate that the magnetoelectric transport can be switched by controlling the ferrielectric domain. The device is cooled through the structural phase transition at 145 K under different gate voltages, and the same measurements as in Fig. 2 are repeated (the resulting four polarization states are summarized in Figs. 4a and 4b, which we derive below). Figures 4c and 4d show the $V_G$ dependence of the graphene resistance under different in-plane $B$ for the same W-BN sample after cooling at $V_G$ = 0 V (Fig. 4c) and at $V_G$ = 4.5 V (Fig. 4d). The measurement sequence is Figs. 2a and 2c (Scan 1), followed by Figs. 4d and 4f (Scan 2), and then Figs. 4c and 4e (Scan 3). The reproducibility between Scans 1 and 3 demonstrates that the behavior under different cooling conditions is fully reversible. Note that the data in Figs. 2a, c (Scan 1) and Figs. 4c, e (Scan 3) are obtained at different in-plane magnetic field angles, B∠b = 140° and 50°, respectively, showing again no significant dependence on the in-plane *B* orientation (see SI Sections 4 and 10).

The key observation is the following: the direction of the hysteresis is unchanged between the two cooling conditions (Figs. 4c and 4d), but the relative magnitude of the *B*-induced CNP shift is swapped between the two CNP peaks (Figs. 4e and 4f). That is, after 4.5 V cooling, the large-ME state is realized on the opposite side of the polarization loop, so that the sign of the polarization of the large-ME state is reversed. We label the two polar states realized after 4.5 V cooling FiE1' and FiE2', such that FiE1' is the state with the larger ME effect. Based on these observations, we now summarize the experimental facts established so far and deduce from them the detailed nature of the four polarization states. The experimental facts are as follows. (1) There are two gate-switchable surface polar states with finite and opposite polarizations (FiE1 and FiE2). (2) A bulk-uniform polarization flip is inconsistent with the observations, and the polarization reversal is strongly suggested to be confined to the near-surface region (see SI Section 5). (3) FiE1 and FiE2 (FiE1' and FiE2') are inequivalent, with FiE1 (FiE1') exhibiting the larger ME response. (4) The sign of the polarization of the larger-ME state is reversed by field cooling, $P_{\mathrm{FiE1}} > 0$ whereas $P_{\mathrm{FiE1'}} < 0$. (5) The (FiE1', FiE2') pair is favored over the (FiE1, FiE2) pair under positive-field cooling. These facts lead uniquely to the four-state picture illustrated in Figs. 4a and 4b. Since the interchange between the two switching pairs occurs only upon cooling through the structural phase transition, the difference between the pairs must be tied to the $CrS_6$–$P_2S_6$ framework, which is distorted at this transition. The (FiE1, FiE2) and (FiE1', FiE2') pairs are realized in the two framework domains. Because the magnitudes of the ME effect of FiE1 and FiE1' (FiE2 and FiE2') are almost the same, FiE1 (FiE2) and FiE1' (FiE2') are related by the space-inversion operation. The polarization can accordingly be decomposed into the dominant, gate-switchable contribution of the $Cu^+$ displacements ($P_{\mathrm{Cu}}$) and a smaller framework contribution ($P_{\mathrm{frame}}$) whose sign is fixed within each framework domain (see SI Section 6 for the detailed schematic). This four-state picture consistently accounts for all of the observations above. Most importantly, it establishes that the electric field cooling reversibly selects the multiferroic domain — and thereby the magnetoelectric transport.

The selection of the framework domain during cooling can be biased by a small out-of-plane electric field. Under 0 V cooling, the selection is governed by a sample-specific built-in field, whereas cooling under $V_{\mathrm{G}}$ = 4.5 V overrides this internal field and reverses the preferred domain (see SI Sections 9 and 17). Because only such a small bias field is required, the domain configuration can be rewritten repeatedly by cooling under different gate voltages, as demonstrated by the full reproducibility between Scans 1 and 3 (Figs. 4c-f). The same field-cooling selection of the framework domain is also confirmed in a WO-BN device by Hall and quantum-oscillation measurements (see SI Section 9), showing that this control is not specific to the device structure. This rewritability distinguishes the interfacial magnetoelectric transport from a fixed, sample-specific response and adds a practical functionality to van der Waals multiferroic interfaces.

## Conclusion

In conclusion, we have demonstrated magnetoelectric transport at a van der Waals heterointerface of graphene and the multiferroic CCPS. The graphene resistance near the CNP is highly sensitive to interfacial carrier-density modulation, directly reflecting the polarization states of CCPS. The resistance peak at the CNP exhibits a pronounced hysteresis, providing clear evidence of polarization switching in CCPS. Quantitative analysis of the CNP peak separation further allows the polarization magnitude to be extracted, confirming consistency with the expected values for the FiE1 and FiE2 states.

Moreover, the CNP position is governed by the spin configuration in CCPS and shifts systematically under applied in-plane magnetic fields, revealing that graphene magnetotransport is directly controlled by the magnetoelectric response of the multiferroic layer. This magnetic-field-dependent modulation is strongly suppressed above the magnetic ordering temperature, evidencing its intrinsic multiferroic origin. We also identify both $B$-even and $B$-odd contributions to the CNP shift, corresponding to second-order and linear ME effects, respectively.

Finally, by cooling the device under different gate voltages, we achieve reversible control of the multiferroic domain configuration, thereby interchanging the FiE1 ⇔ FiE2 switching pair with the FiE1' ⇔ FiE2' switching pair. This further extends the functionality of magnetoelectric transport at vdW interfaces. Our results thus establish the first demonstration of gate-switchable electric transport that directly reflects the magnetoelectric effect. The demonstrated approach is broadly applicable to a variety of van der Waals heterostructures and offers a design principle for functional devices utilizing the magnetoelectric effect, including magnetoelectric spintronic devices based on atomically thin materials with clean interfaces.

## Methods

1. Sample fabrication

The gate/CCPS/ (hBN/) graphene/hBN devices are fabricated entirely within a nitrogen-filled glove box to prevent degradation of the material and ensure a clean interface (See SI Section 2). The back gate and bottom contact electrodes for graphene are patterned using electron beam lithography, followed by the sputtering of a Ti/Au (4/8 nm). Thin flakes of CCPS are mechanically exfoliated from bulk single crystals and transferred onto the pre-patterned back gate using a dry transfer method based on a polycarbonate (PC)/polydimethylsiloxane (PDMS) stamp. Subsequently, a hBN flake with a thickness of approximately 30 nm and a monolayer graphene (and a thin hBN for the W-BN device) are sequentially picked up and released onto the CCPS flake and the bottom contact. For sample S3 (see SI Section 8 for the list of samples), edge contacts to graphene are employed instead of bottom contacts. In this case, reactive ion etching (RIE) is performed with $Ar/CF_4/O_2$ gas, followed by Ti/Au (5/100 nm) sputtering at a finite tilt angle and rotation to form conformal contacts. Finally, to improve spatial uniformity and minimize the effects of interfacial strain or trapped bubbles introduced during the stacking process, the graphene is further patterned into a narrow geometry using RIE with $Ar/CF_4$ gas. Thickness of the CCPS flake is confirmed by atomic force microscopy after the transport measurements.

2. Transport measurement

The devices are wire-bonded with aluminum wires for transport measurements. Transport measurements are carried out using lock-in amplifiers (SRS SR830) in combination with a current preamplifier (SRS SR570). A constant current of 50 nA is applied during all electrical measurements, and back-gate voltages are supplied by a source meter (Keithley 2612). Hysteresis curves are obtained by waiting more than 1.1 s after changing the gate voltage to eliminate hysteresis originating from measurement lag. All measurements are performed in a physical property measurement system (PPMS, Quantum Design).

## Acknowledgements

We acknowledge helpful discussion with Shuichi Iwakiri. M. T. acknowledges support from the JSPS KAKENHI (grant no. 23K19026, 25K17326), Murata Science and Education Foundation, JST, PRESTO (grant no. JPMJPR24H7), and Kondo Memorial Foundation. S. A. acknowledges support from the JSPS KAKENHI (Grant Number JP24KJ0840). K. W. and T. T. acknowledge support from the JSPS KAKENHI (grant no. 21H05233 and 23H02052) , the CREST (JPMJCR24A5), JST and World Premier International Research Center Initiative (WPI), MEXT, Japan. T. Id acknowledges support from the JSPS KAKENHI (Grant Numbers JP23H00088, JP24H01176, JP25H02117) and JST FOREST (Grant Number JPMJFR213A).

**Author contributions**

M. T. conceived and designed the research work, fabricated the device, performed the transport measurement, and analyzed the data. S. A. developed the device fabrication method and participated in the device fabrication. S. A. and N. O. performed the SHG measurements. H. O. and J. P. performed the PFM measurement. M. T., I. S., I. P., T. Is, and M. H. performed the capacitance bridge measurements. M. T., I. S., I. P., T. Is, C. Y., N. T., and M. Y. constructed the measurement setup. T. T. and K. W. grew the hBN crystal. M. T. and T. Id. led the manuscript writing. T. Id. supervised the project. All authors discussed the results and commented on the manuscript.

## Figures

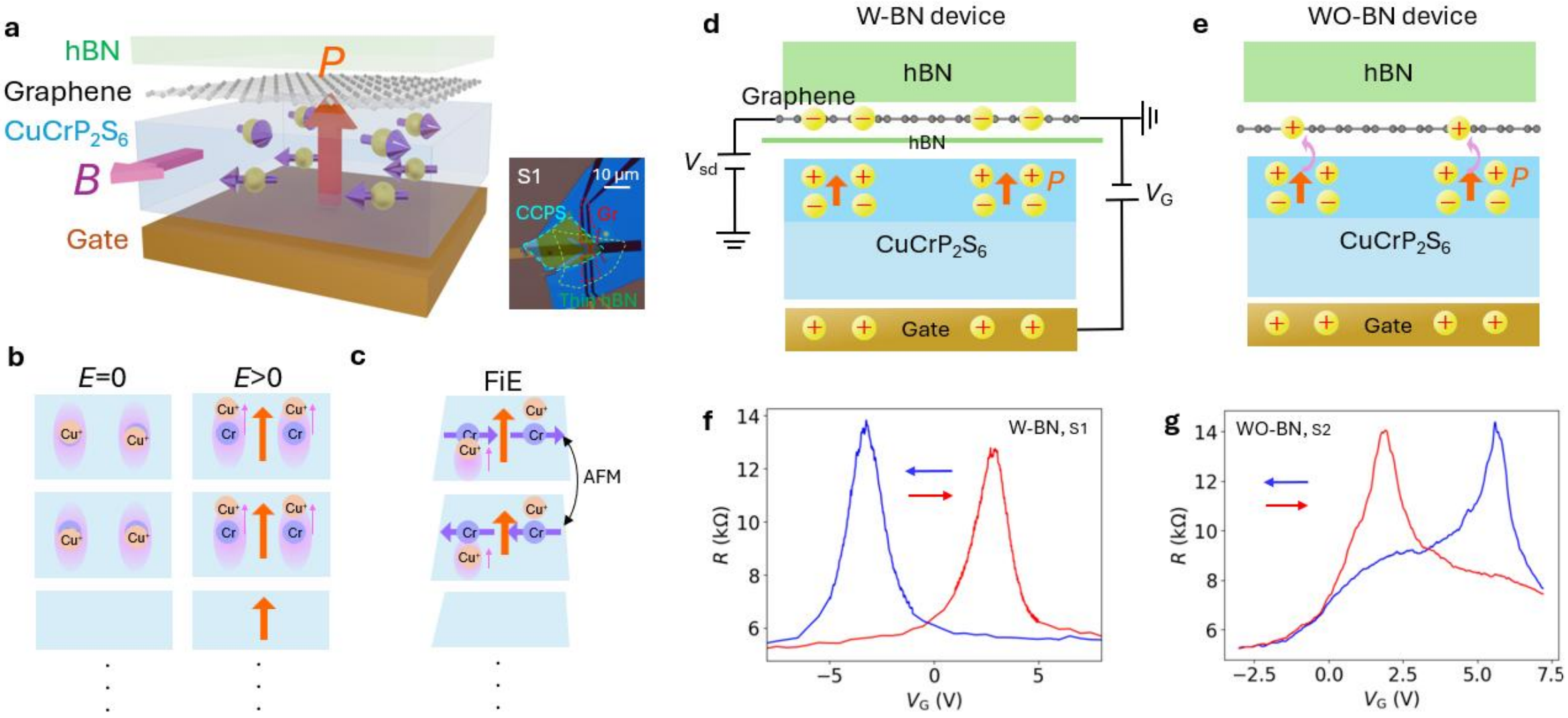

Figure 1: Transport signatures of switchable polarization in graphene/$CuCrP_2S_6$ (CCPS) interfaces.

(a) Schematic of the device structure and magnetoelectric (ME) effect in the CCPS/graphene interface. The carrier density in graphene is modulated by the gate through the CCPS flake, reflecting the polarization states of CCPS. When an in-plane magnetic field is applied, the spin configuration of CCPS is altered, leading to a change in its polarization—i.e., a ME effect. The inset shows an optical microscope image of sample 1 (W-BN device), consisting of an hBN/graphene/6~8 nm thick hBN/CCPS stack on a Ti/Au back gate. Electrical contacts to graphene are made using a bottom electrode underneath the graphene. (b) Schematic illustration of the $Cu^+$ ion and dielectric properties of CCPS above 190 K. In the ground state with zero net polarization (left), $Cu^+$ ions are randomly distributed along the out-of-plane direction around centrosymmetric positions. In the positively polarized state (right), $Cu^+$ ions are displaced toward the top surface. (c) Schematic illustration of the $Cu^+$ ion configurations of the top surface in the low-temperature Pc phase. The bulk ground state is thought to be the so-called AFE state, in which neighboring $Cu^+$ ions are displaced alternately in opposite out-of-plane directions. Because the $CrS_6$–$P_2S_6$ framework breaks inversion symmetry, the cancellation of their dipole moments is not exact. In the surface configuration shown here, one of the $Cu^+$ ions shifts, producing a ferrielectric state with a larger net polarization. The gate-switchable surface states FiE1 and FiE2 carry finite net polarizations of opposite sign. (d, e) Schematic of the two device structures. In the W-BN device (d), a 6~8 nm thick hBN spacer is inserted between graphene and CCPS, while in WO-BN device (e), CCPS is directly attached to graphene. The thickness of CCPS is in the range of 22~49 nm (see SI Section 8 for the thickness of each sample). The blue and light-blue regions indicate the surface and the bulk states, respectively, although in reality the two are continuously connected. $V_G$ is applied between graphene and back gate, with graphene kept at 0 V. A bias voltage for graphene transport measurement is applied between two bottom contacts for graphene across the gated region. Red symbols + and - with yellow circles indicate the accumulated charges. (f, g) Gate voltage dependence of the graphene resistance for the W-BN device (sample 1, f) and the WO-BN device (sample 2, g) at $T$=2 K. The red (blue) curve corresponds to gate sweeps from negative to positive (positive to negative) $V_G$.

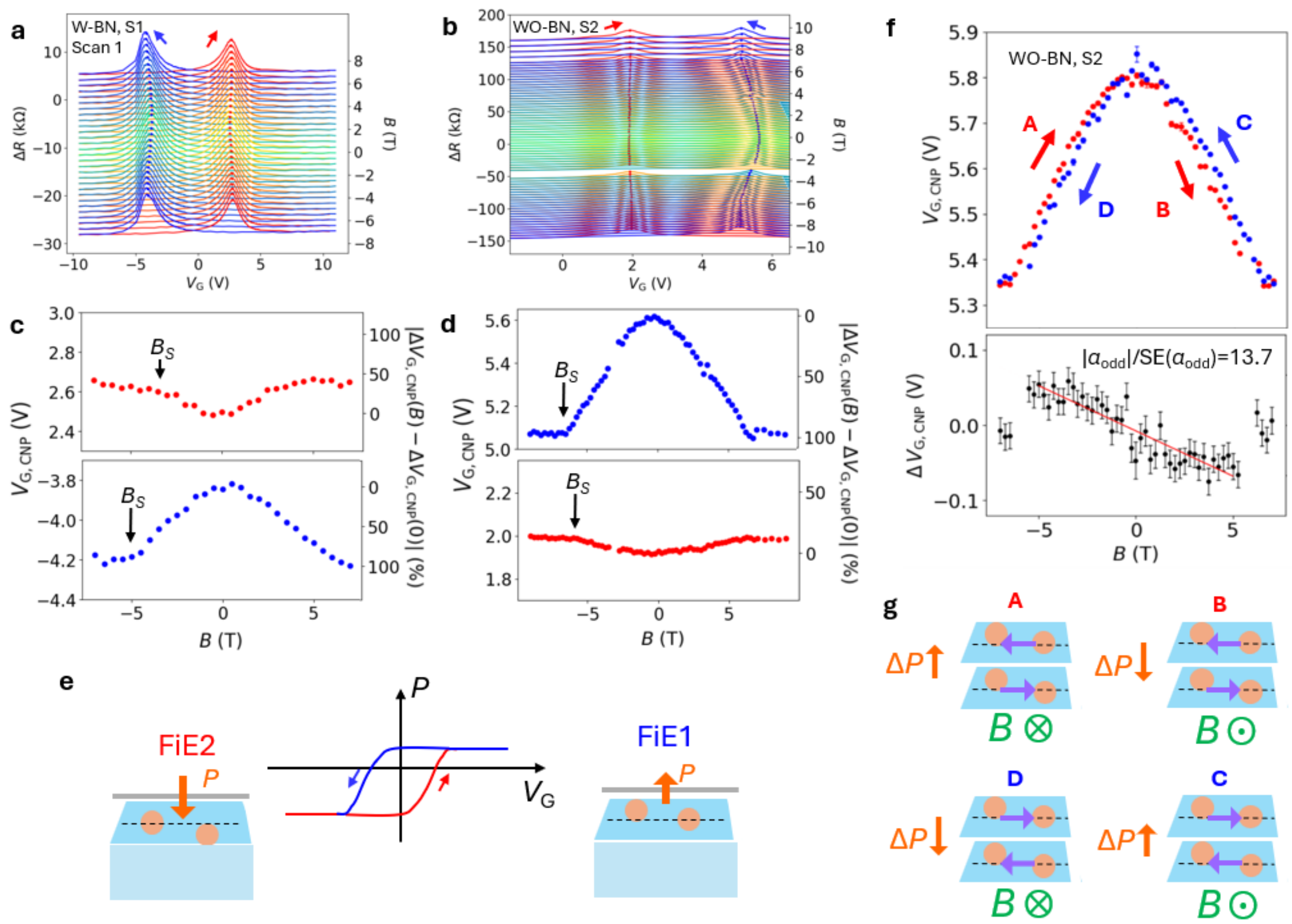

Figure 2: Magnetoelectric transport in graphene/CCPS interfaces.

(a, b) $V_G$ dependence of graphene resistance under different in-plane magnetic fields $B$ in the W-BN (a) and WO-BN (b) devices at $T$=2 K. Curves for different $B$ are vertically offset for clarity, with the field values indicated on the right axis. Red (blue) curves correspond to gate sweeps from negative to positive (positive to negative) $V_G$. The magnetic field direction is $B\angle$b=140° for (a) and $B\angle$b=40° for (b). (c, d) In-plane $B$ dependence of the CNP peak voltage $V_{G,\,CNP}$ in the W-BN (c) and WO-BN (d). $V_{G,\,CNP}$ is obtained by Gaussian fitting to $R(V_G)$ in (a) and (b) near the peak maximum. Top (bottom) panels correspond to CNP peaks located at larger (smaller) $V_G$. The red and blue symbols indicate the gate-sweep direction in which the corresponding CNP peak appears, as denoted by the arrows in (a) and (b). $B_s$ denotes the saturation magnetic field, above which $V_{G,\,CNP}$ becomes nearly independent of $B$. A slow, linear-in-time shift of $V_{G,\,CNP}$ on the timescale of several hours is subtracted as background from the W-BN device data (see SI Section 14 for detail). The right axis shows $|V_{G,\,CNP}(B) - V_{G,\,CNP}(0)|$ normalized to its maximum value in the FiE1 state. (e) Schematic of the $V_G$-dependent polarization and the two distinct surface ferrielectric states, FiE1 and FiE2. After applying a positive $V_G$, the polarization state is FiE1 (blue curve); decreasing $V_G$ reverses the sign of the surface polarization, driving the system into the FiE2 state (red curve). This switching between the FiE1 and FiE2 states is confirmed by detailed $V_G$-dependent measurements with varying maximum $V_G$ (see SI Section 15). (f) Top panel: $V_{G, CNP}$ of the FiE1 state in the WO-BN device as a function of the in-plane $B$ at $T$ = 2 K. Red and blue symbols indicate opposite sweep directions of $B$. A-D denote the four sweep segments: (A) $B$=-7→0 T, (B) $B$=0→+7 T, (C) $B$=+7→0 T, (D) $B$=0→-7 T. The bottom panel is the subtraction between red and blue data in the top panel. $\alpha_{odd}$ is the slope of linear ME effect, and SE($\alpha_{odd}$) is its standard error obtained from a weighted least-squares fit. A signal-to-noise ratio $|\alpha_{odd}|/SE(\alpha_{odd}) \geq 2$ is used as the threshold for a statistically significant slope. (g) Schematic of the sign of the linear ME effect for the four configurations of $B$ and the Néel vector, corresponding to the sweep segments A-D in (f).

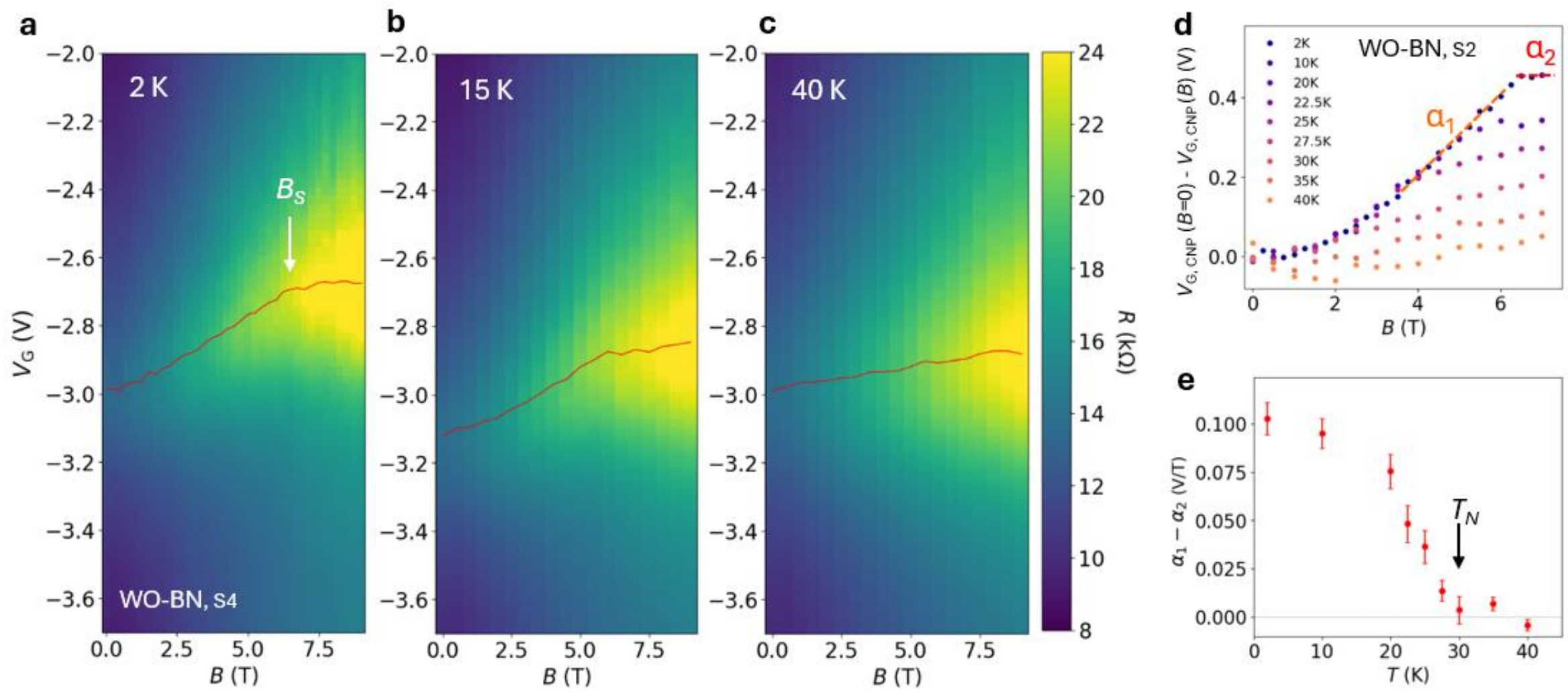

Figure 3: Temperature dependence of magnetoelectric transport.

(a-c) Graphene resistance $R$ as a function of in-plane magnetic field $B$ and $V_G$ in WO-BN device (sample 4) at $T$=2 K (a), 15 K (b), and 40 K (c). Red curves indicate $V_{G,\,CNP}$ obtained from Gaussian fitting to $R(V_G)$ near the peak maximum. $B_s$ denotes the saturation magnetic field, above which $V_{G,\,CNP}$ becomes nearly independent of $B$. At 40 K, which is above the Néel temperature $T_N$, saturation behavior of $V_{G,\,CNP}$ ($B$) is not observed. (d) $V_{G,\,CNP}(B=0) - V_{G,\,CNP}(B)$ as a function of in-plane $B$ in the WO-BN device (sample 2) at different temperatures. $\alpha_1$ and $\alpha_2$ indicate the slopes $\frac{\partial \Delta V_{G,CNP}}{\partial B}$ obtained from linear fitting in the magnetic field range of 3.5 T < $B$ < 6 T ($\alpha_1$) and 6 T < $B$ ($\alpha_2$), respectively. (e) Temperature dependence of $\alpha_1 - \alpha_2$, which characterizes the saturation behavior of $V_{G,\,CNP}$ ($B$). Error bars represent the standard errors of the linear fits. The black arrow indicates the Néel temperature $T_N$.

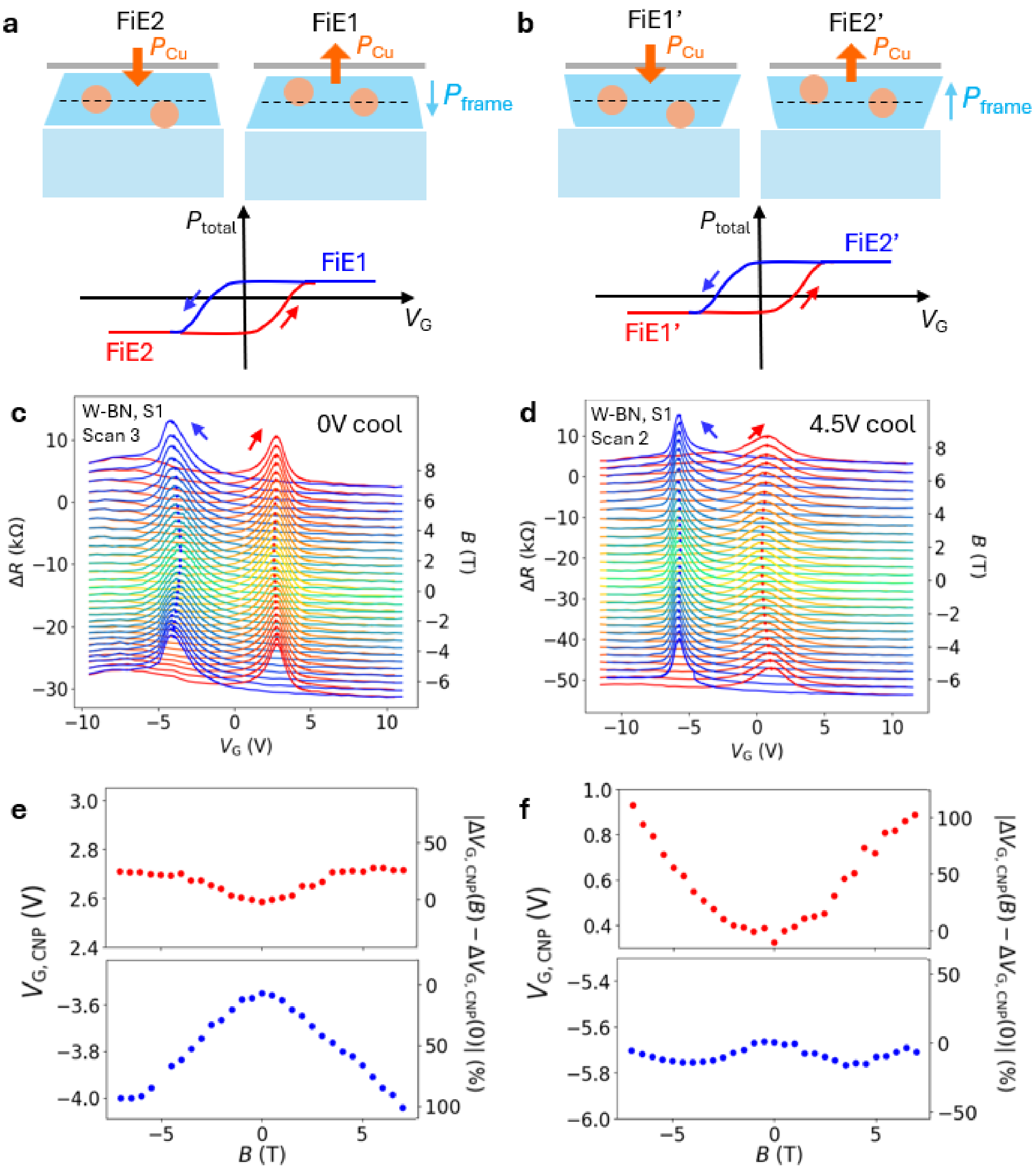

Figure 4: Controlling multiferroic domains and switchable magnetoelectric transport.

(a, b) Schematic illustration of the four distinguishable polarization states and the corresponding $P$–$V_G$ hysteresis loops. (a) The (FiE1, FiE2) pair: FiE1 has a positive and FiE2 a negative net surface polarization. (b) The (FiE1', FiE2') pair, realized after cooling under a finite gate voltage: FiE2' has a positive and FiE1' a negative net surface polarization. In each schematic, the pink circles represent the $Cu^{+}$ ions in the near-surface layer, and the distorted trapezoid represents the $CrS_6$–$P_2S_6$ framework, whose distortion is opposite between the two pairs (see SI Section 6). Red and blue curves indicate the polarization as a function of $V_G$, where the arrows correspond to the $V_G$ sweep directions indicated in (c) and (d). (c, d) $V_G$ dependence of the graphene resistance at $T$=2 K under different in-plane magnetic fields $B$ in the W-BN device (sample 1) for two cooling conditions: $V_G$=0 V (c) and $V_G$=4.5 V (d). The data in (c) and (e) are obtained from a different scan than that shown in Fig. 2a, c, under a different in-plane magnetic field angle $B\angle$b=50°. The results confirm the reproducibility of the $B$-dependence behavior and show no significant dependence on the in-plane $B$ orientation. Red and blue arrows indicate the $V_G$ sweep direction. (e, f) In-plane $B$ dependence of the CNP peak voltage $V_{G,\,CNP}$ after $V_G$=0 V cooling (e) and $V_G$=4.5 V cooling (f). $V_{G,\,CNP}$ is obtained by Gaussian fitting to $R(V_G)$ in (c) and (d) near the peak maximum. Top (bottom) panels correspond to CNP peaks located at larger (smaller) $V_G$. The red and blue symbols indicate the gate-sweep direction in which the corresponding CNP peak appears, as denoted by the arrows in (c) and (d). A slow, linear-in-time shift of $V_{G,\,CNP}$ on the timescale of several hours is subtracted as background (see SI Section 14 for detail). The right axis represents $|V_{G,CNP}(B) - V_{G,CNP}(0)|$ normalized to its maximum value in the FiE1 (FiE1') state.

Supplementary Information for

# Switchable Magnetoelectric Transport in Graphene via a Van der Waals Multiferroic

Miuko Tanaka*[1], Shunta Aoki[1], Ikoi Sato[1], Hao Ou[2], Itishree Pradhan[1], Ngoc Han Tu[5],

Yangsong Chen[1], Tomohiro Ishii[1], Kenji Watanabe[3], Takashi Taniguchi[4],

Michihisa Yamamoto[5,6], Masayuki Hashisaka[1], Jiang Pu[2], Naoki Ogawa[5], Toshiya Ideue*[1]

*[1]Institute for Solid State Physics, The University of Tokyo, Kashiwa-shi, Japan*

*[2] Department of Physics, Institute of Science Tokyo, Meguro-ku, Japan*

*[3]Research Center for Electronic and Optical Materials,*

*National Institute for Materials Science, Tsukuba, Japan*

*[4]Research Center for Materials Nanoarchitectonics,*

*National Institute for Materials Science, Tsukuba, Japan*

*[5]Center for Emergent Matter Science, RIKEN, Wako-shi, Japan*

*[6]Quantum-Phase Electronics Center and Department of Applied Physics,*

*The University of Tokyo, Tokyo, Japan*

*Corresponding authors

## 1. Crystal structure of $CuCrP_2S_6$

Figure S1 shows schematic illustrations of the crystal and magnetic structures of $CuCrP_2S_6$ (CCPS). CCPS is a layered material composed of a honeycomb lattice of distorted $CrS_6$ octahedra, $CuS_3$ triangles, and pairs of *P* ions inside the honeycomb network. At $T > 190$ K, mobile $Cu^+$ ions occupy either the upper or lower side of the van der Waals layer with equal probability, leading to a centrosymmetric structure with space group *C2/c* (Figs. S1a, b). In this phase, ionic conduction behavior has been reported [23, 44-46]. Although the zero-electric-field ground state does not break inversion symmetry, it exhibits hysteretic *P-E* curve owing to slow ionic dynamics against the external electric field [22-24]. Upon cooling, the motion of $Cu^+$ ions starts to freeze at $T = 190$ K and completely settles down at $T = 145$ K, yielding a structural transition to a non-centrosymmetric phase with space group *Pc* (Figs. S1c, d). With further decrease in temperature, CCPS exhibits antiferromagnetic order below the Néel temperature $T_N = 32$ K, in which spins of $Cr^{3+}$ ions align along the a-axis (Fig. S1e). In this magnetically ordered state, both spatial inversion symmetry and time-reversal symmetry are broken. The easy-axis anisotropy along the a-axis is approximately 0.3 T; therefore, the Néel vector aligns perpendicular to the applied in-plane magnetic field above 0.3 T, which is consistent with the experimental observation that the magnetoelectric transport shows no significant dependence on the *B* direction as discussed in Section 10.

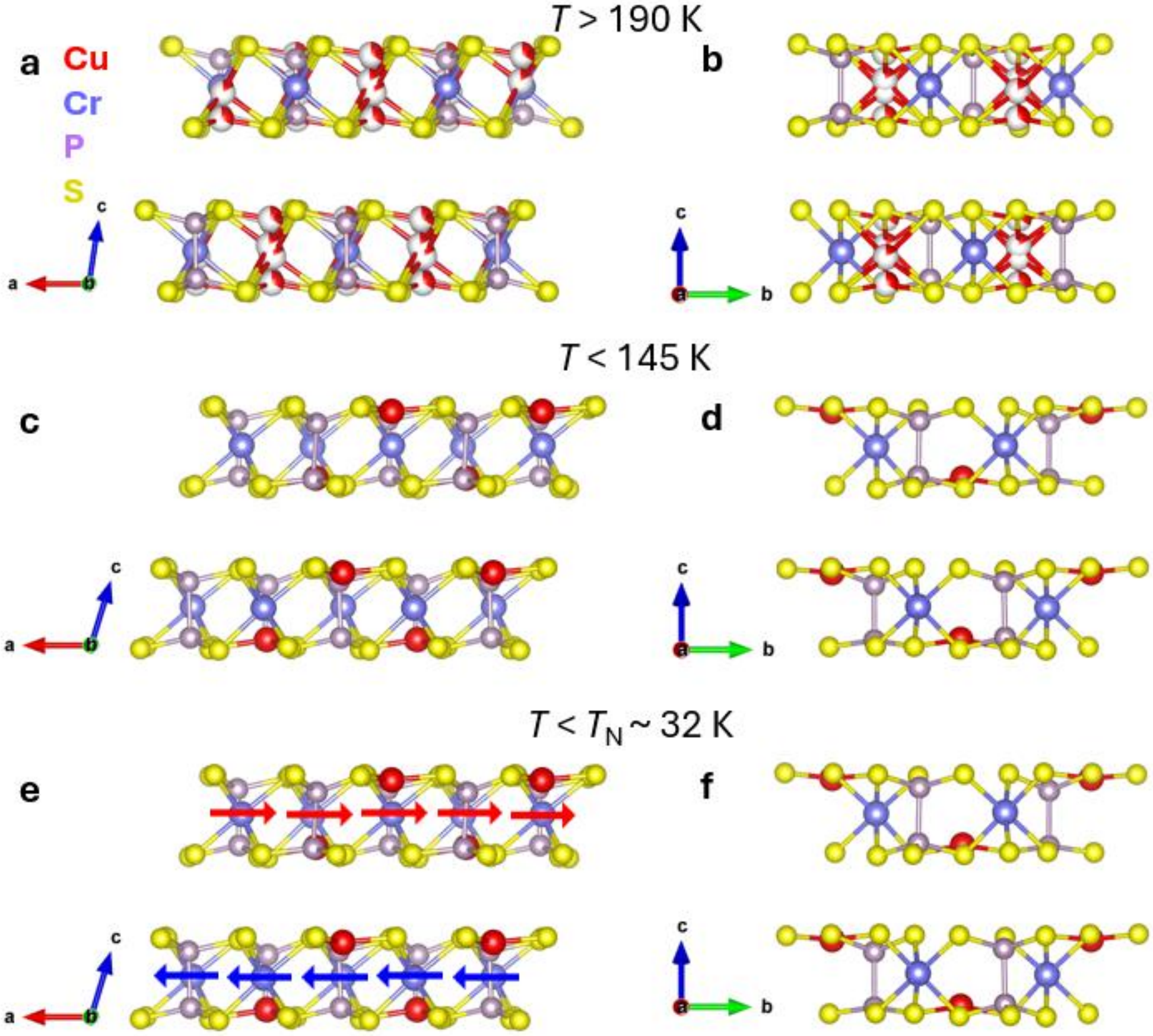

Figure S1: Crystal structures and magnetic properties of $CuCrP_2S_6$ (CCPS).

(a, b) Crystal structures of CCPS above 190 K viewed along the b-axis (a) and a-axis (b). In this phase, mobile $Cu^+$ ions are randomly distributed on either side of the 2D plane, giving a centrosymmetric structure with space group *C2/c*. Schematics were created using VESTA 3 [50]. (c, d) Crystal structures of CCPS below 145 K viewed along the b-axis (c) and a-axis (d). In this low-temperature phase, $Cu^+$ ions are ordered, resulting in a non-centrosymmetric structure with space group *Pc*. (e, f) Magnetic structure of CCPS below $T_N$ = 32 K, showing A-type antiferromagnetic order with $Cr^{3+}$ spins aligned along the a-axis.

## 2. Second harmonic generation to determine crystal axis

Optical second harmonic generation (SHG) is employed to determine the crystal axis. Sample substrates are placed on a pillar-shaped copper mount with varnish and inserted into a cryostat (Cryo Industries). Fundamental laser pulses at 1.55 e*V* photon energy are generated using a Ti: Sapphire amplifier (1 kHz, 100 fs) and sent to the sample with pseudo-Köhler geometry, and the SHG images at 3.1 e*V* reflected from the sample are detected by a cooled CCD camera (PIXIS: 1024B). The linear polarizations of the incident light and the SHG signals are set to be parallel. For the polarization dependence measurements, the half-wave plate is rotated from 0° to 90°.

Figure S4 shows the SHG signal at 300 K, 50 K, and 5 K. At 300 K, where the inversion symmetry is preserved, small SHG signal, possibly from surface or electric quadrupole contribution, is observed (Fig. S4a). After the structural phase transition at $T$ = 145 K, CCPS becomes non-centrosymmetric and the electric dipole term of the SHG is allowed, which generates the SHG along the a-axis (Fig. S4b). In the multiferroic phase below $T$ = 32 K, another SHG contribution appears along the b-axis (Fig. S4c), which can be interpreted as SHG signals originating from the magnetic dipole term [28].

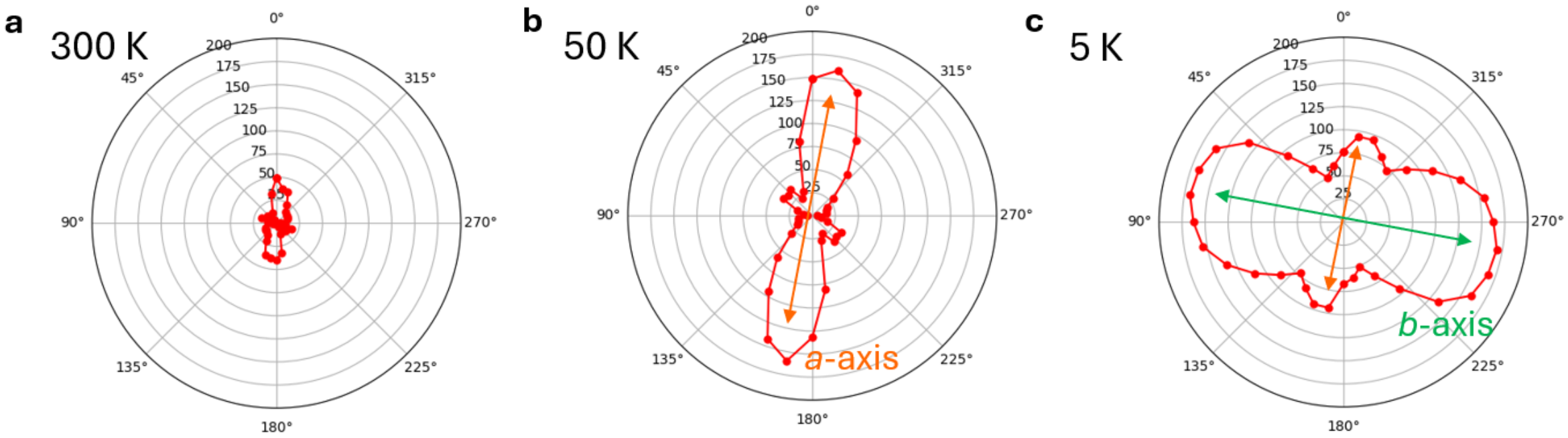

Figure S4: Polarization dependence of SHG signals at 300 K (a, paramagnetic phase with space group *C2/c*), 50 K (b, paramagnetic phase with space group *Pc*), and $T$ = 5 K (c, antiferromagnetic phase).

## 3. Capacitance measurements and the surface polarization flip

To evaluate the bulk and surface contributions to the ferrielectric behavior, we employ a parallel-plate capacitor (PPC) with the structure [Au (50 nm) / CCPS (d = 44 nm) / bilayer graphene] (S8). The stack is fabricated on a GaAs substrate to minimize parasitic capacitance to approximately 0.8 fF, compared with a much larger value (~7 pF) for a $Si/SiO_2$ substrate. Two complementary probes are applied to the same device: the carrier density of the bilayer-graphene electrode extracted from quantum-capacitance oscillations (sensitive to the surface bound charge) and the total capacitance between Au and bilayer graphene across the CCPS flake (averaging the dielectric response over the CCPS thickness).

We first note that the electrostatic doping of graphene, sensed either by the graphene resistance or by the quantum capacitance, is determined by the polarization-induced bound charge, and is insensitive to how the polarization is distributed across the thickness (see Section 13 for the electrostatic model); the graphene carrier density alone therefore cannot distinguish a bulk-wide from a surface-localized polarization reversal, which is why the bulk-averaging capacitance measurement is essential.

The capacitance is measured using an AC capacitance bridge with an AC excitation voltage of 0.05 V at a frequency of 10 kHz (Fig. S6a). The bilayer graphene is kept at 0 V and a DC voltage $V_{DC}$ is applied to the bottom Au electrode; the capacitance is measured as a function of $V_{DC}$ and of the out-of-plane magnetic field.

The capacitance is around 0.52 pF (Fig. S6b), which corresponds to $\varepsilon_{CCPS}/\varepsilon_0 = 7.9$ considering the thickness and the area of the PPC, in reasonable agreement with $\varepsilon_{CCPS}/\varepsilon_0 = 7.7$ obtained from another device (S10) described at the end of this section. The total capacitance of the device is the series combination of the geometric capacitance $C_{geom}$ set by the CCPS dielectric constant and the quantum capacitance $C_Q$ of graphene, $1/C_{total} = 1/C_{geom} + 1/C_Q$ (Fig. S6a). The quantum capacitance is proportional to the density of states ρ of the graphene. This contribution to the total capacitance is usually negligible in metal–insulator–metal PPCs because ρ of a metal is very large. In our device, because bilayer graphene is used as one of the electrodes, this effect is observable [55]. In particular, at $T = 2$ K under a perpendicular magnetic field, the Landau quantization leads to a periodic $\rho(E_F)$ (Fig. S6a, right panel), which is observed as an oscillating capacitance as a function of $V_{DC}$ (Fig. S6c). Note that this oscillation owing to the quantum capacitance is only ~2 % of the total capacitance (Fig. S6b, c).

One period of this oscillation corresponds to a change of the carrier density by $n_{LL}$ = 4eB/h, which is the degeneracy of a Landau level (4 accounts for the spin and valley degeneracy in bilayer graphene). Using this relation, the oscillation is converted to the relative carrier density of graphene as a function of $V_{DC}$ (Fig. S6c, bottom panels). Within each sweep direction, the slope of $\Delta n_{Gr}(V_{DC})$ agrees with that expected from the geometric capacitance of the CCPS layer with $\varepsilon_{CCPS}/\varepsilon_0$ = 7.7~7.9 (dashed lines in Fig. S6c), confirming the quantitative consistency of this conversion. A clear anti-hysteresis is observed, which is consistent with the anti-hysteretic graphene resistance as a function of gate voltage shown in the main text (Fig. 1g). This observation confirms that the polarization flip of CCPS occurs in this device consistently with the devices presented in the main text.

If the entire bulk contributed to the polarization flip, there should be a significant hysteretic $V_{DC}$ dependence of $\varepsilon_{CCPS}$, because the change of the polarization is reflected in the dielectric constant through $\varepsilon = dP/dE$. Although only a fraction of the polarization change across the flipping process is reversible and reflected in the small-signal AC measurement, a change of the dielectric constant of typically tens of percent to more than an order of magnitude is observed at the polarization flip in ferroelectrics [71-74].

However, the observed capacitance $C_{total}$ as a function of $V_{DC}$ shows totally different behavior (Fig. S6b). It does not show any peaks, and the hysteretic variation of $C_{total}$ with $V_{DC}$ remains below 0.5 % (Fig. S6b, inset). This strongly indicates that a bulk-uniform polarization flip does not occur in this system: it would require a dielectric response one to two orders of magnitude smaller than any of the values reported above. This is also consistent with the theoretical calculation [26], in which the electric field required for the polarization flip lies outside the accessible range for bulk CCPS, while it decreases rapidly with decreasing flipping layer number.

These results can be explained if the polarization reversal is confined to a thin surface layer; the device capacitance is the series combination of a non-switching bulk capacitance and a thin switching capacitance, and the total capacitance remains essentially $V_{DC}$-independent. In this case, the total capacitance is expressed as

$$\frac{1}{C_{\mathrm{total}}} = \frac{d_{\mathrm{CCPS}} - d_{\mathrm{surface}}}{\varepsilon_{\mathrm{CCPS}} S} + \frac{d_{\mathrm{surface}}}{(\varepsilon_{\mathrm{CCPS}} + \Delta\varepsilon_{\mathrm{CCPS}}) S} + \frac{1}{C_{\mathrm{Q}}}$$

Here, we assume an abrupt boundary between the polarization-flipping surface layer and the non-switching interior (in reality the crossover should be gradual), and $d_{surface}$ denotes the thickness of the polarization-flipping region. The relative change of the total capacitance at the flip is $\Delta C_{total}/C_{total} \approx (d_{surface}/d_{CCPS}) \times \Delta\varepsilon_{CCPS}/(\varepsilon_{CCPS} + \Delta\varepsilon_{CCPS})$. For example, for the most conservative reported value $\Delta\varepsilon/\varepsilon \approx 10\%$, the observed hysteretic variation of $C_{total}$ with $V_{DC}$ below 0.5 % therefore sets an upper bound $d_{surface} < 0.005\, d_{CCPS}\, (\varepsilon_{CCPS} + \Delta\varepsilon_{CCPS})/\Delta\varepsilon_{CCPS} \approx 2.4$ nm, corresponding to 4 layers (the monolayer thickness is ≈ 0.64 nm); for larger typical values of $\Delta\varepsilon/\varepsilon$ the upper bound becomes thinner.

Thus, the coexistence of the two observations in the same sample — a clearly hysteretic surface polarization seen by the graphene carrier density, and a $V_{DC}$-independent bulk-averaged dielectric constant — is inconsistent with a bulk-uniform polarization flip and strongly suggests that the polarization reversal in CCPS is confined to the near-surface region. This surface-localized ferrielectric behavior on top of a non-switching interior is analogous to reported examples in other materials, including $PbZrO_3$ [56–58] and layered chalcogenides such as GeSe and SnTe [59, 60]. Incomplete screening of the depolarization field at surfaces and interfaces is also known to modify polar configurations relative to the bulk [63, 64]. A similar surface-versus-bulk dichotomy is known in the magnetism of $CrI_3$: even bulk crystals host surface layers whose stacking — and hence magnetic order — differs from the interior [48, 49], as verified spectroscopically by magneto-Raman measurements [65].

To further strengthen our conclusion, we also examined whether inhomogeneity of the polarization other than the bulk–surface contrast could explain the above experimental observations.

1. In-plane inhomogeneity

The possibility of an in-plane inhomogeneous coexistence of switched and unswitched regions is constrained by the single CNP peak of graphene. If the spread of switching voltages among coexisting regions were larger than the typical CNP width, multiple peaks should appear in the graphene resistance (Fig. S5a). The observed single peak therefore bounds the inhomogeneity of the polarization flip to below the CNP width. The polarization-induced CNP displacement is several times larger than the CNP width, so such small inhomogeneity does not affect the conclusion. Therefore, in-plane inhomogeneity cannot explain the combination of the clear polarization flip probed by the graphene doping and the almost flat dielectric response.

An alternative scenario may be that some regions never switch while others do. In that case, two CNPs should appear, one with hysteresis and one without (Fig. S5b). Such a double-peak structure is not observed in our data.

In fact, in samples where the CCPS flake was mistakenly folded during stacking and contained wrinkles or cracks, multi-peak structures arising from this kind of in-plane inhomogeneity are observed (Fig. S5c). All such multi-peak samples were excluded, and the discussion in the paper is based only on samples that are uniform within the CNP peak width.

2. Out-of-plane inhomogeneity

If the polarization flip is inhomogeneous in the thickness direction, the change of the dielectric constant is reduced by the ratio of the thickness of the flipping region to the total thickness, and this can explain the observed flat capacitance. The limiting case in which the flipping layer is confined to the surface is exactly the surface–bulk picture proposed here.

The measurements do not uniquely determine the depth profile of the polarization, and the possibility that the flipping layer lies inside the bulk cannot be directly excluded. However, there is no reason for a few specific layers to switch selectively inside the translationally invariant interior, and even if translational symmetry were broken there by some extrinsic cause, it is difficult to see how such a coincidence would reproduce across all our samples. The surface, where translational symmetry is intrinsically broken, is therefore by far the most reasonable location of the switching layers.

3. Screening by a conductive internal layer

A sufficiently conductive internal region would modify the electrostatic boundary conditions and effectively decouple part of the CCPS thickness. The internal conductive region would be electrically floating because it is not connected to an electrode. We consider the simplest limiting case of the conductive layer in the middle of CCPS which is homogeneous in the in-plane direction (Fig. S5d). In this case, the flake behaves as a bulk-uniform system with a reduced effective thickness. Therefore, this possibility is excluded in the same way as the bulk-uniform polarization flip scenario. If this conductive region existed inhomogeneously in the plane, it would result in the multi-peak behavior already argued to be unlikely in the in-plane discussion above.

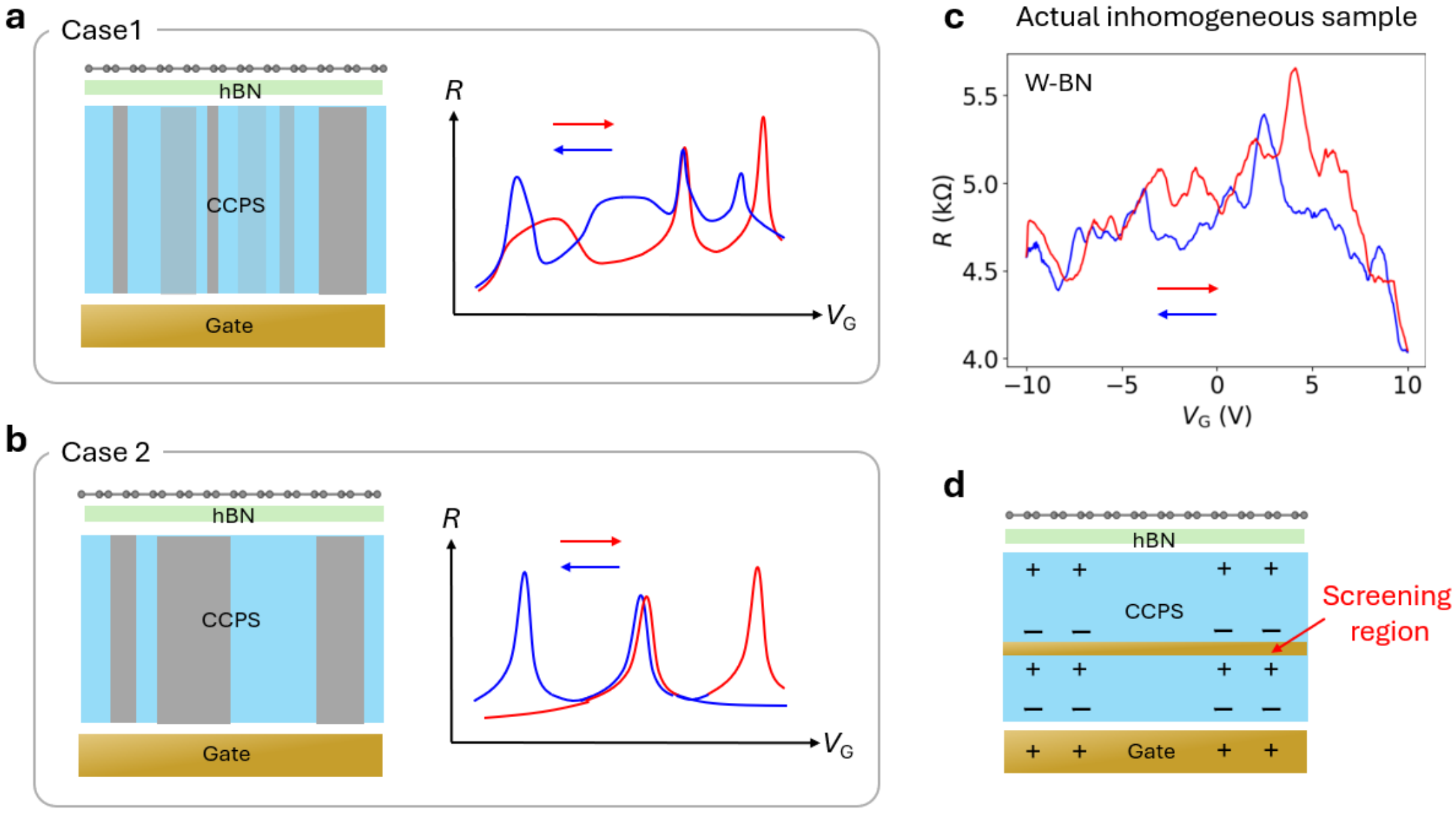

Figure S5: Alternative scenarios for the polarization flip. (a) Case 1, where regions with substantially different switching voltages coexist, producing multiple hysteretic CNP peaks. (b) Case 2, where switched and unswitched regions coexist, producing two CNP peaks, one with and one without hysteresis. (c) Gate-voltage dependence of the resistance measured in an inhomogeneous W-BN sample containing wrinkles and cracks, showing a multi-peak structure. Such samples were excluded from the analysis. (d) CCPS containing an internal conductive region that is electrically floating. The flake then behaves as if its effective thickness were reduced, with the graphene doping and the dielectric response unchanged from a uniform bulk.

The same device also allows us to examine the out-of-plane magnetic-field dependence of the dielectric constant of CCPS. As shown in Fig. S6b, the capacitance at $B$ = 7 T applied perpendicular to the plane coincides with that at $B$ = 0 T within the measurement accuracy: apart from the quantum-capacitance oscillation of the graphene electrode (~2 % of the total capacitance), no significant out-of-plane field dependence of $\varepsilon_{CCPS}$ is observed.

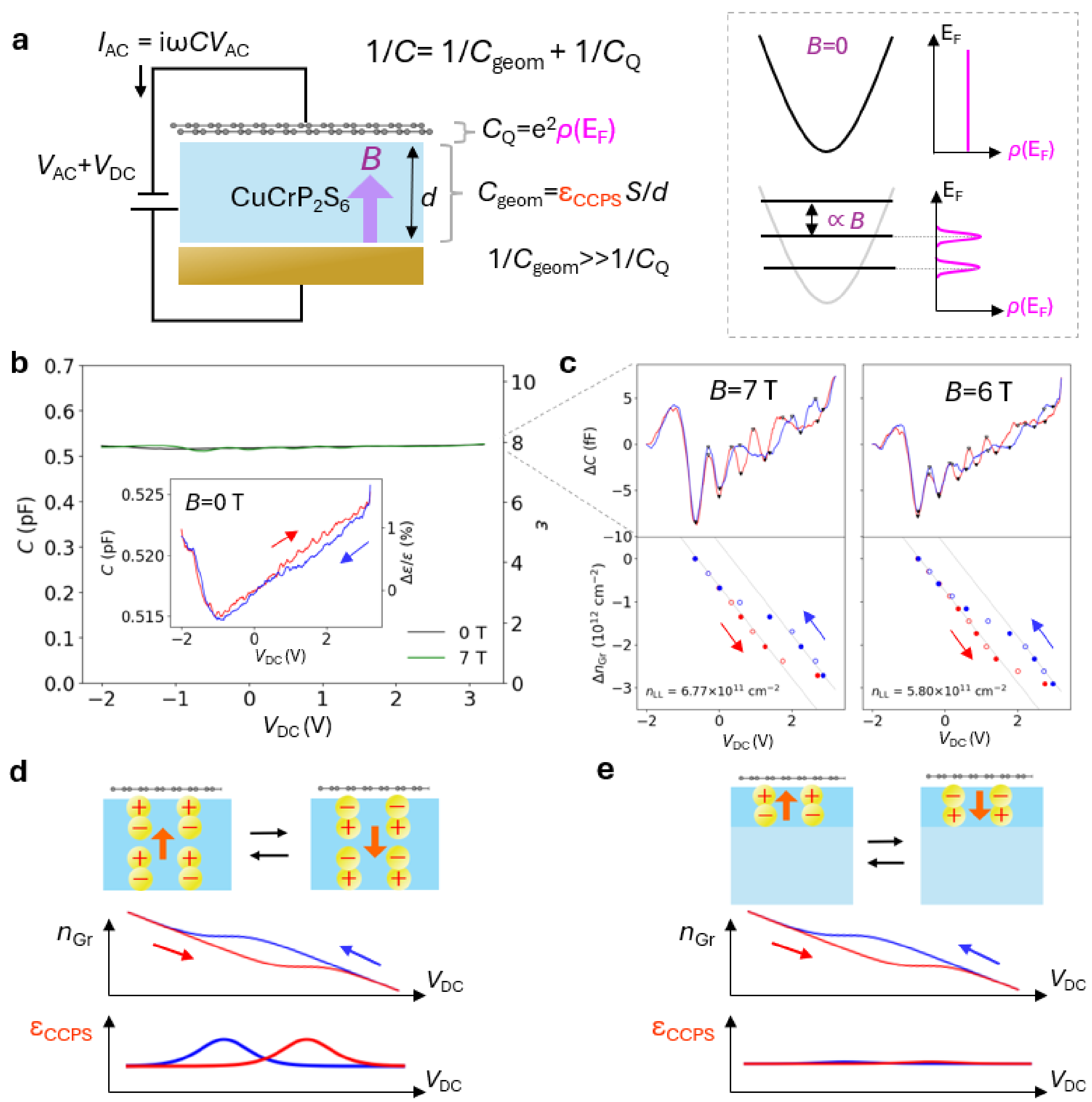

Figure S6: Capacitance measurements on an Au/CCPS/bilayer-graphene parallel-plate capacitor (PPC). (a) Left: side-view schematic of the device and the measurement circuit. An AC excitation

$V_{AC}$ = 0.05 V superimposed on a DC bias $V_{DC}$ is applied, and the capacitance is obtained from the AC current $I_{AC}$ = iωCVAC. The magnetic field is applied out of the plane. The total capacitance is the series combination of the geometric capacitance $C_{geom} = \varepsilon_{CCPS}$ S/d and the quantum capacitance $C_Q = e^2\rho(E_F)$ of the bilayer graphene ($1/C_{geom} >> 1/C_Q$). Right: schematic of the density of states $\rho(E_F)$ of bilayer graphene without and with an out-of-plane magnetic field. Under Landau quantization, $\rho(E_F)$ oscillates as $E_F$ is tuned, which appears as a small oscillation of the capacitance. (b) Capacitance as a function of $V_{DC}$ at $B$ = 0 T (grey) and 7 T (green); the right axis shows the corresponding $\varepsilon_{CCPS}$. Inset: magnified view at $B$ = 0 T. Red and blue traces correspond to opposite $V_{DC}$ sweep directions, and the right axis shows Δε/ε (%). No peak structure is observed, and the total variation remains below ~2 %, and the difference between the two sweep directions (the hysteretic component) remains below ~0.5 %. (c) Top panels: oscillatory component of the capacitance ΔC as a function of $V_{DC}$ at $B$ = 7 T and 6 T. Red and blue traces correspond to opposite $V_{DC}$ sweep directions; black dots mark the oscillation extrema. Bottom panels: relative carrier density of graphene $\Delta n_{Gr}$ obtained by counting the oscillation periods in the top panels, where one period corresponds to $n_{LL}$ = 4eB/h ($6.77 \times 10^{11}$ cm$^{-2}$ at 7 T and $5.80 \times 10^{11}$ cm$^{-2}$ at 6 T). A negative (positive) sign of the carrier density corresponds to electron (hole) doping. A clear anti-hysteresis between the two sweep directions is observed, consistent with the anti-hysteretic transfer curve of the WO-BN devices (main text Fig. 1g). Dashed lines show the $V_{DC}$ dependence of the carrier density expected from the geometric capacitance with $\varepsilon_{CCPS}/\varepsilon_0$ = 7.7. (d) Schematic of the limiting case in which the entire CCPS bulk contributes to the polarization reversal (top), and the expected $V_{DC}$ dependence of the graphene carrier density $n_{Gr}$ and of $\varepsilon_{CCPS}$ (bottom). If the whole bulk switched, $\varepsilon_{CCPS}$ should exhibit a hysteretic peak near the polarization flip. (e) The same schematics for the case in which only a thin near-surface layer reverses, as inferred for CCPS in the present study: $n_{Gr}$ shows the same hysteresis, while $\varepsilon_{CCPS}$ remains featureless, in agreement with the observation in (b).

The qualitatively same behavior is reproduced in another device (S9) with the structure [graphite (5–7 nm) / CCPS (45 nm) / 5–6-layer graphene], whose data are shown in Fig. S7. Quantum oscillations are again observed in the magnetic-field dependence of the capacitance, measured here as a function of 1/$B$ at fixed $V_G$ ($T$ = 10 K, Fig. S7a) [55].

The oscillation period depends on $V_G$. Using the degeneracy of the Landau level

$$n = 4e / (h \cdot \Delta(1/B)),$$

with the spin–valley degeneracy g = 4 of graphene, the carrier density extracted from the oscillation period is shown in Fig. S7b. The blue points correspond to a protocol in which $V_G$ = +9.5 V is applied first and then swept to +5 V and 0 V; the red points are obtained after first applying −9.5 V and then sweeping to −5 V and 0 V. Within each sweep direction, the $V_G$ dependence of $n$ follows the relation expected from a relative dielectric constant of 7.7 (dashed lines), and the offset between the two branches at $V_G$ = 0 corresponds to a hysteresis of the surface polarization. The total capacitance, on the other hand, shows no peak as a function of $V_G$, with a variation below 0.6 % (Fig. S7c), fully consistent with the behavior of S8.

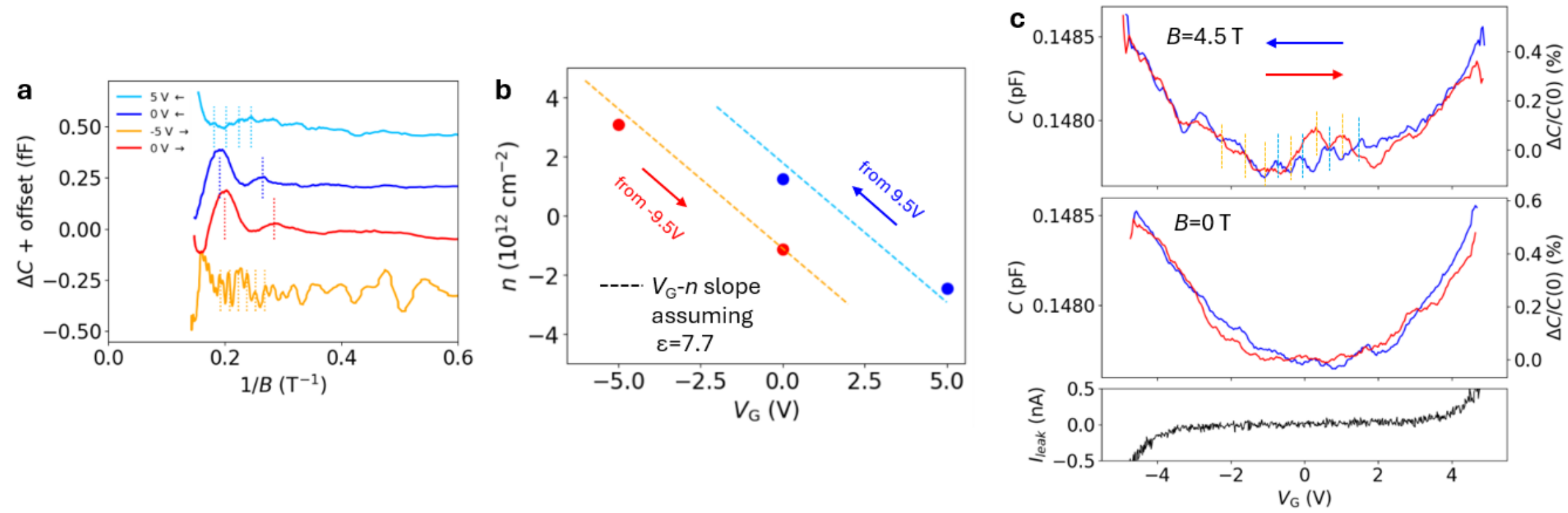

Figure S7: Capacitance measurements on the graphite(5-7 nm)/CCPS/multilayer graphene (5-6 layer) parallel-plate capacitor (S9). (a) Oscillatory component of the capacitance ΔC as a function of $1/B$ at $T$ = 10 K for different gate voltages (curves are vertically offset for clarity; dashed ticks mark the oscillation maxima). The blue and red traces at $V_G$ = 0 V are taken after initialization at $V_G$ = +9.5 V and −9.5 V, respectively. (b) Carrier density extracted from the oscillation periods in (a). A negative (positive) carrier density corresponds to electron (hole) doping, following the sign convention of Fig. S6. Dashed lines show the $V_G$–$n$ relation expected from a CCPS relative dielectric constant of 7.7. Since the quantum oscillations in (a) determine only the magnitude of the carrier density, the sign of the $V_G$ = 0 V data points is assigned so that they align with the dashed line corresponding to the dielectric constant of 7.7. (c) Capacitance as a function of $V_G$ at $B$ = 4.5 T (top) and $B$ = 0 T (middle); the right axes show ΔC/C(0) (%). No peak structure is observed, and the variation remains below 0.6 %. The bottom panel shows the leakage current between the two electrodes.

In addition, in-plane magnetic-field dependence is measured in another parallel-plate capacitor, in which a CCPS flake is sandwiched between two Ti/Au electrodes (S10) (Figs. S8a and S8b). The same excitation voltage of 0.05 V and the frequency of 10 kHz are used for the measurement. Figure S8c shows the temperature dependence of the capacitance, which exhibits a kink at the structural phase transition. No significant kink is observed at the magnetic phase transition, indicating a negligible influence of magnetism on the dielectric constant. The lowest-temperature value of the capacitance corresponds to a relative dielectric constant of 7.7, consistently with S8 and S9. This value is smaller than the previously reported bulk value of approximately 12. No significant magnetic field dependence exceeding 2% is observed, suggesting that the observed $B$ dependence of $V_{G, CNP}$ presented in the main text originates not from the $B$ dependence of dielectric constant but from the $B$ dependence of the spontaneous polarization.

An extremely large $B$ dependence of the dielectric constant has been reported in bulk CCPS, reaching approximately -99% relative to the value at $B$=0 T [43], which appears inconsistent with our measurements in CCPS nanoflakes. Furthermore, whereas no significant second-order $B$-even ME effect has been observed in bulk CCPS, we detect a second-order contribution that exceeds the linear ME effect. This discrepancy may arise from the presence of mixed ferrielectric domains in bulk crystals, or from possible differences in stacking order between bulk and nanoflakes, similar to observations in the van der Waals magnet $CrI_3$ [48, 49].

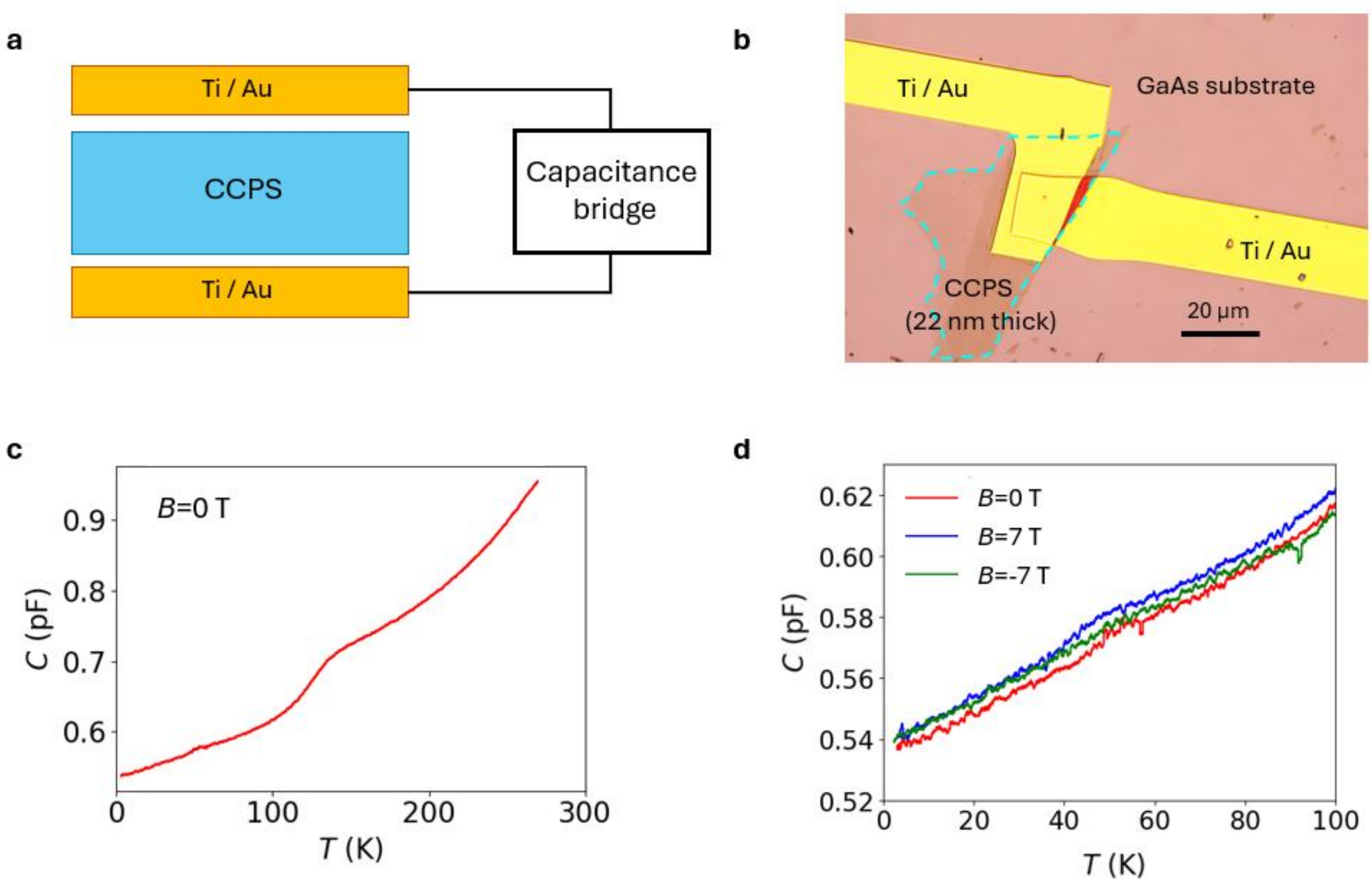

Figure S8: Dielectric constant measurement in a parallel plate capacitor sample.

(a) Schematic cross section of the parallel-plate capacitor sample and the capacitance bridge measurement setup. (b) Optical microscope image of the parallel-plate capacitor sample. (c, d) Capacitance as a function of temperature under in-plane magnetic field of $B$=0 T (c, red curve in d), $B$=7 T (blue curve in d), and $B$=-7 T (green curve in d).

## 4. Summary of measured samples and reproducibility in other samples

We examined 14 samples S1–S14 with transport measurement, as summarized in Table S1. Samples for PFM and SHG measurement are not listed here.

S1, S6, and S7 are W-BN devices, which contain a thin hBN spacer (6–8 nm, 11–12 nm, and 9–10 nm, respectively) between CCPS and graphene, whereas S2–S4, S8, 9, S12-14 are WO-BN devices, in which CCPS is directly attached to graphene. S5 contains both regions: an hBN spacer (12 nm) is inserted under the left half of the graphene, while the right half directly contacts CCPS (see Section 9). The CCPS thickness is comparable among the samples, ranging from 22 nm to 49 nm (except for tunneling samples). The back gate of S6 does not work, and that of S5 failed after the small-range gate sweeps shown in Fig. S15d; these two samples are used for the Hall measurement (Section 9). S8–S10 are used for AC capacitance measurement (Section 5). S11 is bilayer graphene encapsulated by hBN and dual gate, where the extrinsic offset doping level of graphene/hBN interface is examined (Section 9). S12 and S13 are tunneling devices with thin CCPS, where the effect of Au/CCPS interface is examined (Section 13). In S14, the Hall effect and quantum oscillations are measured under a perpendicular magnetic field to examine the carrier density and the field-cooling effect (Section 9).

Each gated sample realizes one of the two switching pairs introduced in Section 6 — (FiE1, FiE2) or (FiE1', FiE2') — depending on the $CrS_6$–$P_2S_6$ framework domain (fourth column of Table S1). The pair is assigned from the sign of the relative polarization of the state with the larger ME response, following the criterion of Section 6. In S1 and S14, as shown in Fig. 4 of the main text and SI section 9 for S1 and S14, respectively, the framework domain — and hence the realized pair — can be reversed by cooling under a finite gate voltage. In S1, 0 V cooling gives (FiE1, FiE2), while 4.5 V cooling gives (FiE1', FiE2'). In S14, 0 V cooling gives (FiE1, FiE2), while 4 V cooling gives (FiE1', FiE2'). In S7, the (FiE1', FiE2') pair is realized already under 0 V cooling; that is, the framework domain is opposite to that of S1. For S2–S4, although intentional control by field cooling is not performed, the naturally realized pairs differ among the samples: S2 and S3 realize (FiE1, FiE2), whereas S4 realizes (FiE1', FiE2') (Table S1, Fig. S14c–f). As discussed in Section 17, this variation is likely due to differences in residual or environmental electric fields — arising from fabrication details, charged impurities, and structural

variations — that bias the selection of the framework domain at the structural phase transition.

Despite these differences in the framework domain, the magnitude of the polarization difference between the two states of a pair — quantified by $|V_{G,\,CNP,\,FiE1} - V_{G,\,CNP,\,FiE2}|$ (fifth column of Table S1; the values in parentheses are the corresponding electric-field scales) — is comparable across samples with the same device structure. This indicates that, while the framework domain may vary between devices, the intrinsic magnitude of the polarization difference and the magnitude of the charge transfer are reproducible for a given device configuration. The value is larger in S1, probably because the charge-transfer effect is absent owing to the hBN spacer, whereas S7 exhibits a value comparable to S2–S4, possibly owing to a thickness dependence. The magnetic-field-induced CNP shifts of the FiE1 (FiE1') peak and of the FiE2 (FiE2') peak (sixth and seventh columns) are comparable among all five gated samples, and the shift of the FiE1 (FiE1') peak is always the larger, consistent with the assignment criterion above. The signs of the *B*-induced shifts are common within samples realizing the same pair and are reversed between the (FiE1, FiE2) and (FiE1', FiE2') samples (colors in Table S1), consistent with the field-cooling reversal shown in Fig. 4 of the main text. Note that the relative CNP positions of the two states are opposite between the W-BN and WO-BN structures (sign of the fifth column), reflecting the opposite hysteresis direction discussed in the main text, whereas the *B*-induced shift has the same sign. These observations also indicate that the magnetic-field dependence of the charge transfer is negligible and that the magnetic-field dependence of the polarization is the dominant contribution for both hBN-inserted and non-inserted samples.

In Figure S14, the $V_G$ dependence of the resistance and the magnetic-field dependence of $V_{G,\,CNP}$ in S3 and S4 are summarized. They show quantitatively consistent behavior compared with S2. The data of S7 are shown in Section 11.

| # | Structure | CCPS thickness | FiE1-FiE2 or FiE1'-FiE2' | $V_{G,CNP,FiE1}$ $-V_{G,CNP,FiE2}$ | $V_{G,CNP,FiE1}$ ($B$=0 T) $-V_{G,CNP,FiE1}$ ($B$=7 T) | $V_{G,CNP,FiE2}$ ($B$=0 T) $-V_{G,CNP,FiE2}$ ($B$=7 T) |
|---|---|---|---|---|---|---|
| S1 | Au/CCPS/ hBN(6~8 nm)/Gr | 39 nm | FiE1-FiE2 *FiE' for 4.5V cooling | -6.0 V (-153 mV/nm) | +0.45 V (11.5 mV/nm) | -0.12 V (3.08 mV/nm) |
| S2 | Au/CCPS/Gr | 42nm | FiE1-FiE2 | +3.82 V (91 mV/nm) | +0.53 V (12.6 mV/nm) | -0.083 V (1.98 mV/nm) |
| S3 | Au/CCPS/Gr | 32nm | FiE1-FiE2 | +2.54 V (79 mV/nm) | +0.35 V (10.9 mV/nm) | -0.051 V (1.56 mV/nm) |
| S4 | Au/CCPS/Gr | 30nm | FiE1'-FiE2' | -2.3 V (-76 mV/nm) | -0.29 V (9.67 mV/nm) | +0.082 V (2.73 mV/nm) |
| S5 | Au/CCPS/ hBN(12 nm)*/Gr | 45 nm | *hBN is under half of graphene | | | |
| S6 | Au/CCPS/ hBN(11-12 nm)/Gr | 26 nm | - | | | |
| S7 | Au/CCPS/ hBN(9-10 nm)/Gr | 22 nm | FiE1'-FiE2' | +2.0 V (91 mV/nm) | -0.33 V (15 mV/nm) | +0.115 V (5.2 mV/nm) |
| S8 | Au/CCPS/2L Gr | 44 nm | Gate and $B_\perp$ dependence of Capacitance | | | |
| S9 | Graphite/CCPS /5-6L Gr | 45 nm | Gate and $B_\perp$ dependence of Capacitance | | | |
| S10 | Au/CCPS/Au | 22 nm | $T$ and $B_\parallel$ dependence of capacitance | | | |
| S11 | Au/hBN/ 2L Gr/hBN/Au | - | Gate dependence | | | |
| S12 | Graphite/CCPS /Graphite | 9 nm | Tunneling transport | | | |
| S13 | Au/CCPS/Graphite | 7 nm | Tunneling transport | | | |
| S14 | Au/CCPS/Gr | 49 nm | Swapped by Field cool | 2.9 V (59 mV/nm) | $B_\perp$ dependence | |

Table S1: Summary of device structures and polarization/magnetoelectric characteristics. "FiE1-FiE2 or FiE1'-FiE2'" denotes which switching pair is realized in the sample; for S1, the pair is selected by the cooling voltage (main text Fig. 4). $V_{G,\ CNP,\ FiE1}$ − $V_{G,\ CNP,\ FiE2}$ quantifies the difference in the CNP peak voltage between the two states of the pair, and the values in parentheses are the corresponding electric-field scales. $V_{G,\ CNP,\ FiE1}$ ($B$=0 T) − $V_{G,\ CNP,\ FiE1}$ ($B$=7 T) and $V_{G,\ CNP,\ FiE2}$ ($B$=0 T) − $V_{G,\ CNP,\ FiE2}$ ($B$=7 T) represent the magnetic-field-induced CNP shifts of the FiE1 (FiE1') and FiE2 (FiE2') peaks, respectively. The colors indicate the signs of the entries.

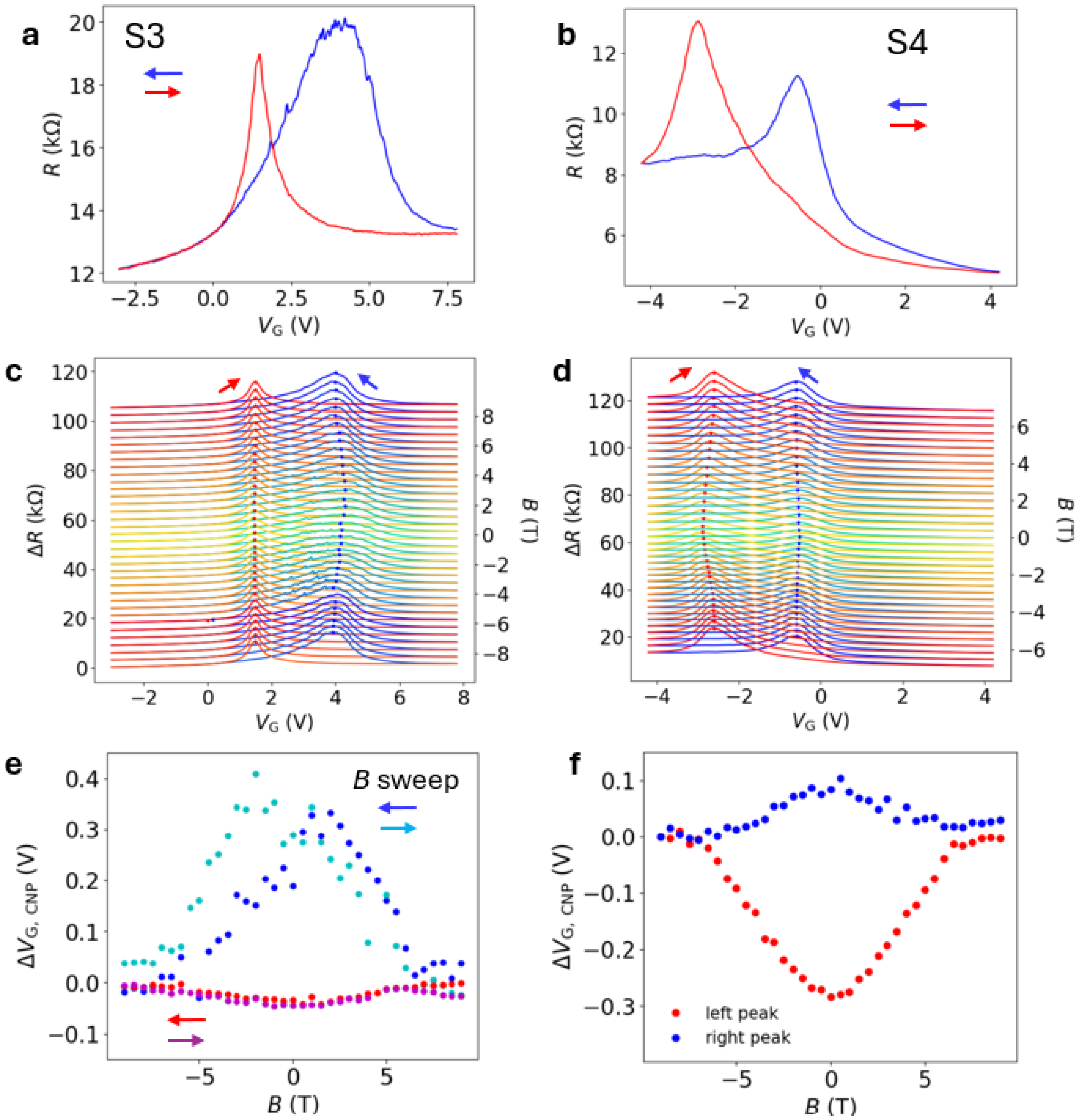

Figure S14: Reproducibility of hysteresis and magnetoelectric response in different samples. (a, b) $V_G$ dependence of graphene resistance for S3 (a) and S4 (b), showing CNP hysteresis between forward (red) and reverse (blue) sweeps. (c, d) $B$ dependence of resistance curves for S3 (c) and S4 (d). (e) $B$ dependence of CNP peak voltage $\Delta V_{G,\,CNP}= V_{G,\,CNP}- V_{G,\,CNP}$ ($B$=7 T) for S3, extracted from (c). (f) $B$ dependence of CNP peak voltage $\Delta V_{G,\,CNP}= V_{G,\,CNP}- V_{G,\,CNP}$ ($B$=7 T) for S4, extracted from (d).

## 5. Measurement time and sweep protocol dependence of $V_{G, CNP}$

Gate sweep rate dependence is examined to identify the origin of hysteresis in the $V_G$ dependence of resistance. Under two different sweep speeds, 28.5 mV/s and 10 mV/s, no significant difference is observed (Fig. S28a). These sweeps were performed in the sequence −4.2 V → 4.2 V → −4.2 V. After the 10 mV/s sweep (blue curve in Figure S28a), the voltage is kept at −4.2 V for 3 minutes and then swept at 28.5 mV/s (red curve in Fig. S28a), again showing no significant difference. This indicates that the origin of hysteresis is not a delayed response but rather a first-order transition of the polarization states.

In ferroelectric FETs, it has been reported that molecular impurities such as water or organic residues at the interface can cause hysteretic behavior due to the delayed response of charge trapping and release processes, which typically occur on the timescale of a few seconds to a few minutes [11,13-15]. The absence of time dependence in the range of a few seconds to a few minutes eliminates this mechanism as the origin of the anti-hysteresis. In addition, the absence of any slower response is confirmed in the sample structure WO-BN. In Fig. S28b, the magnetic field dependence of $V_{G, CNP}$ is shown. After a magnetic field sweep of −9 T → 9 T → −9 T, which takes 6.8 hours, $V_{G, CNP}$ returns to the same value. This confirms that there is no slow charge dynamics even on the timescale of several hours. This behavior is consistently observed in all WO-BN samples.

In contrast, W-BN sample 1, which contains a thin hBN spacer between CCPS and graphene, exhibits a slow drift of $V_{G, CNP}$ on the order of 30 mV/hour. When the magnetic field is swept from −7 T → 7 T → −7 T (Fig. S28c) and from −0.72 T → 0.72 T (Fig. S28e), a shift of the same order is observed as a function of time, even though the magnetic field sweep range differs by a factor of 10. This indicates that a parameter-independent, time-dependent shift of $V_{G, CNP}$ occurs. Since the total measurement time for one cooldown is approximately 30 hours, and $V_{G, CNP}$ is found at slightly different voltages (approximately ±1 V) for each cooldown after thermal cycling, we did not observe saturation of this slow time dependence. This shift is not observed in W-BN device S7, which has a thicker hBN spacer.

A plausible origin of this shift is charge transfer from CCPS to graphene across the 6–8 nm thick hBN layer, where the carrier tunneling rate is quite low due to the relatively large thickness of the hBN. A 30 mV shift of $V_{G, CNP}$ in one hour corresponds to a change in the graphene carrier density of approximately $2.4 \times 10^{14}$ m$^{-2}$. Therefore, the corresponding carrier transfer rate is approximately $1\times10^{-20}$ C/s·μm$^2$. This slow, linear-in-time shift of $V_{G,\,CNP}$ on the timescale of several hours is subtracted as background from the W-BN device data presented in the main text (Fig. S28d).

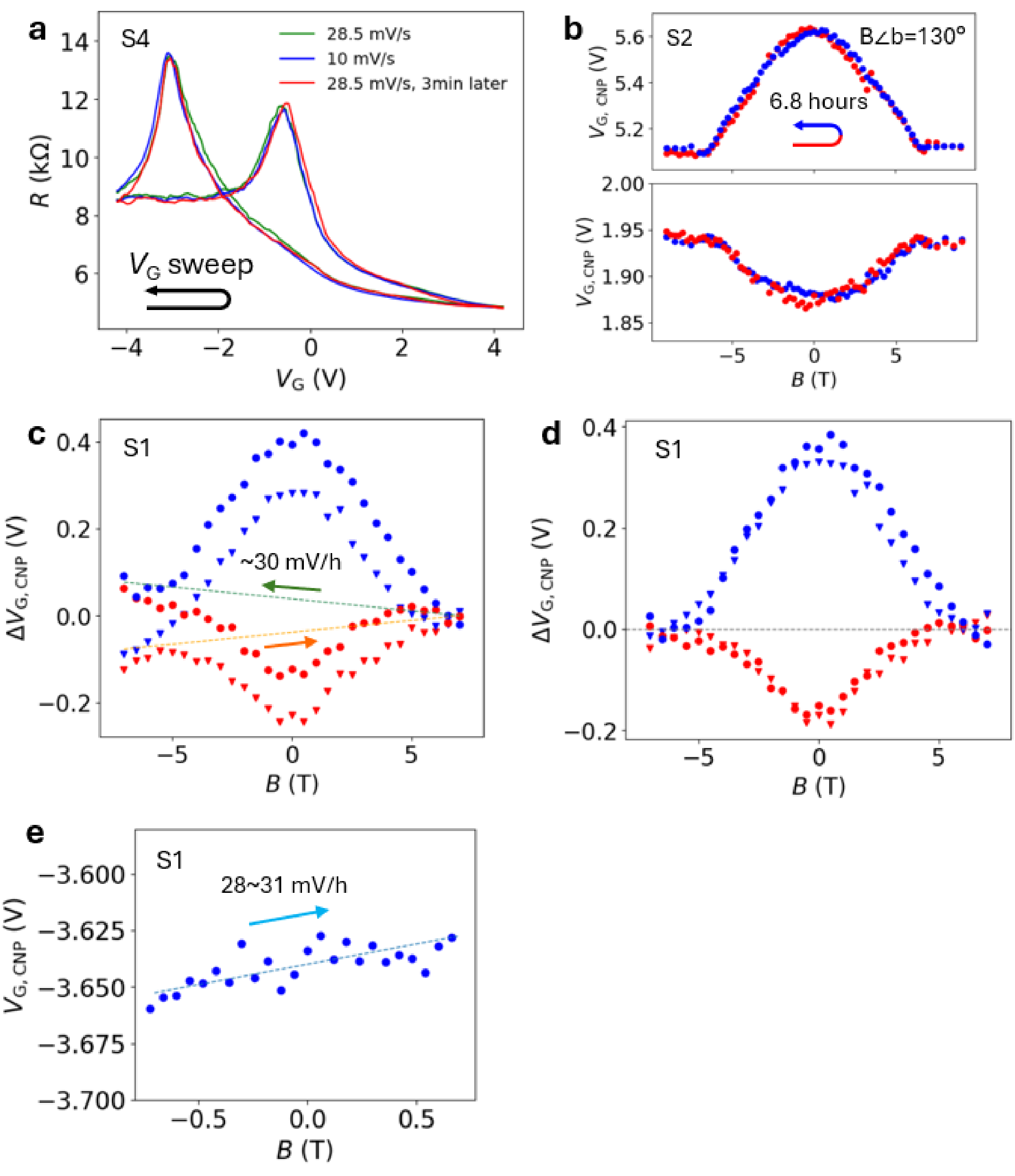

Figure S28: Measurement time dependence of $V_{G,\,CNP}$ .

(a) Resistance of graphene as a function of $V_G$ with different sweeping rates and waiting times at $B$=0 T in sample 4. $V_G$ is swept from −4.2 V → 4.2 V → −4.2 V. The green curve and the blue curve are taken consecutively with an interval of approximately 10 seconds. After the blue-curve sweep, $V_G$ is fixed at −4.2 V and, following a waiting period of approximately 3 minutes, the red curve is taken. (b) $V_{G, CNP}$ of the FiE1 state (top panel) and the FiE2 state (bottom panel) as a function of in-plane magnetic field (b=130°). The magnetic field is swept from −9 T → 9 T → −9 T, which takes approximately 6.8 hours. (c) $\Delta V_{G, CNP}$= $V_{G, CNP}$- $V_{G, CNP}$ ($B$=7 T) of the FiE1 state as a function of in-plane magnetic field (b=130°) in sample 1. The orange and green lines indicate the shift of $V_{G, CNP}$ proportional to time. The magnetic field is swept from −7 T → 7 T → −7 T, which takes approximately 7.1 hours. (d) Data from (c) after subtraction of the time-proportional shift of $\Delta V_{G, CNP}$. (e) $V_{G, CNP}$ of the FiE1 state as a function of in-plane magnetic field (∠b = 130°) in sample 1. The magnetic field is swept from −0.72 T → 0.72 T, which takes approximately 45 minutes.

The transfer curves are independent of the order of the sweeps, including the reversed protocol. Figure S29a shows S1 (W-BN) at fixed magnetic field with four consecutive gate sweeps performed in the order positive→negative (blue), negative→positive (red), positive→negative (cyan dashed), negative→positive (magenta dashed). The curves for the same sweep direction overlap. The same behavior is observed in S2 (WO-BN, Fig. S29b).

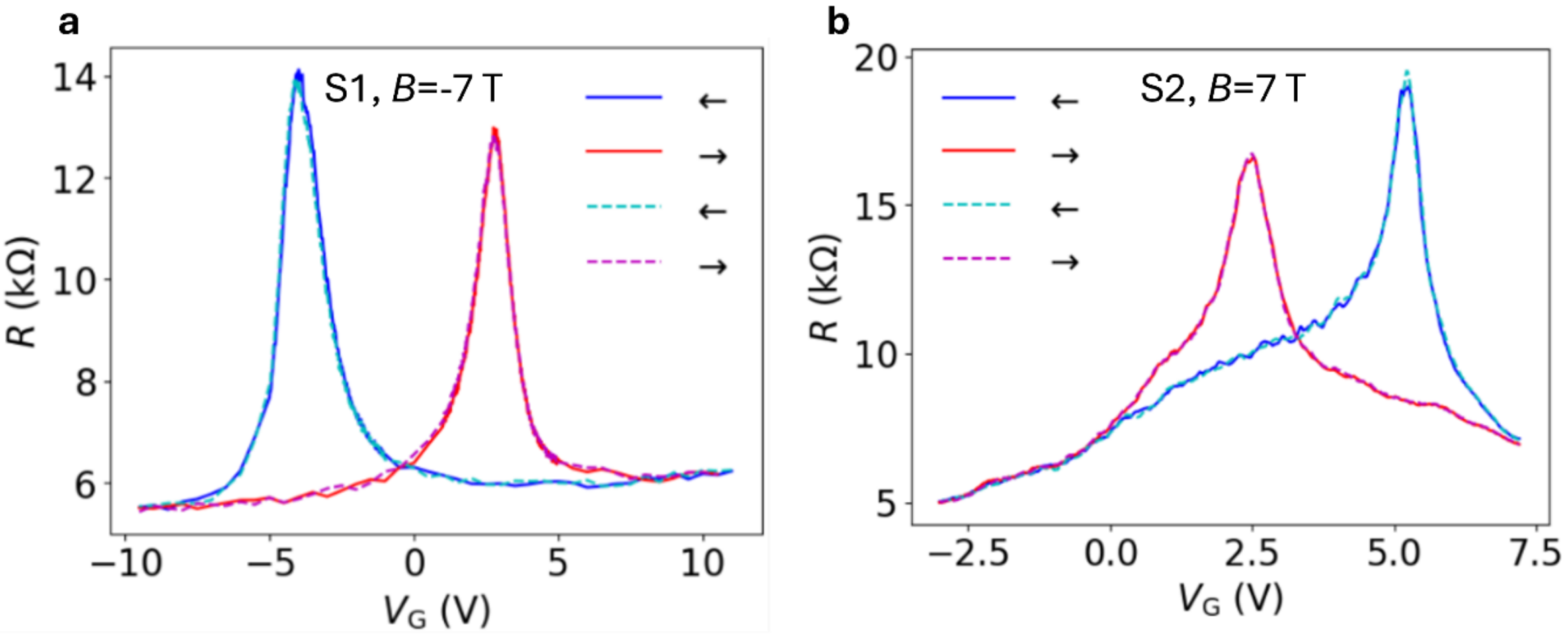

Figure S29: $V_G$ dependence of the graphene resistance in four successive gate sweeps at fixed magnetic field. (a) S1 (W-BN); (b) S2 (WO-BN). The data are taken in the order blue, red, cyan, magenta, with 1–10 s between sweeps.